\documentclass[preprint,notoc]{JHEP3}
\usepackage{graphicx}

\newcommand{\event}{ {\sc event2}}

\newcommand{\Ord}{\mathcal{O}}
 
\newcommand{\nn}{\nonumber \\}
\newcommand{\el}{\nonumber \\ &&}

\newcommand{\eqref}[1]{(\ref{#1})}
\newcommand{\text}[1]{{\rm #1}}
\newenvironment{changemargin}[2]{%
  \begin{list}{}{%
    \setlength{\topsep}{0pt}%
    \setlength{\leftmargin}{#1}%
    \setlength{\rightmargin}{#2}%
    \setlength{\listparindent}{\parindent}%
    \setlength{\itemindent}{\parindent}%
    \setlength{\parsep}{\parskip}%
  }%
  \item[]}{\end{list}}

\newcommand{\be}{\begin{equation}} 
\newcommand{\ee}{\end{equation}} 
\newcommand{\cmb}{\begin{changemargin}}
\newcommand{\cme}{\end{changemargin}}
\newcommand{\bea}{\begin{eqnarray}} 
\newcommand{\eea}{\end{eqnarray}} 
\newcommand{\D}{\delta}

\newcommand{\G}{\Gamma}

\newcommand{\gcusp}{\Gamma_{\rm cusp}}
\newcommand{\rr}{\frac{R}{1-R}}
\newcommand{\lib}{{\mathrm{Li}_2}}

\newcommand{\bn}{{\bar{n}}}

\newcommand{\dd}{\mathrm{d}}
\newcommand{\als}{\alpha_s}
\newcommand{\ep}{\epsilon}

\def\spa#1.#2{\langle#1\,#2\rangle}
\def\spb#1.#2{[#1\,#2]}
\def\sandmm#1.#2.#3{%
\left\langle\smash{#1}{\rphantom1}\right|{#2}%
\left|\smash{#3}{\rphantom1}\right]}
\def\spab#1.#2.#3{\sandmm#1.#2.#3}
\def\spba#1.#2.#3{\sandpp#1.#2.#3}
\def\spaa#1.#2.#3.#4{\sandmp#1.{#2#3}.#4}
\def\spbb#1.#2.#3.#4{\sandpm#1.{#2#3}.#4}
\def\spash#1.#2{\spa{\smash{#1}}.{\smash{#2}}}
\def\spbsh#1.#2{\spb{\smash{#1}}.{\smash{#2}}}

\def\ksl{\not{\hbox{\kern-2.3pt $k$}}}

\def\e{\epsilon}

\def\Ord{{\cal O}}

\def\in{{\rm in}}
\def\out{{\rm out}}
\def\tom{\tau_\omega}

\preprint{IFT-UAM/CSIC-11-96}
\title{Jet mass with a jet veto at two loops and the universality of non-global structure}

\author{Randall Kelley and Matthew D. Schwartz\\
Center for the Fundamental Laws of Nature,\\
Harvard University,\\
Cambridge, MA 02138, USA}
\author{Robert M. Schabinger\\
Instituto de F\'{i}sica Te\'{o}rica UAM/CSIC and \\
Departamento de F\'{i}sica Te\'{o}rica,\\
	Universidad Aut\'{o}noma de Madrid,\\
        Cantoblanco, E-28049 Madrid, Espa\~{n}a}
\author{Hua Xing Zhu\\
Department of Physics and State Key Laboratory of Nuclear Physics and Technology,\\
Peking University,\\
Beijing 100871, China}
        
\abstract{We investigate the exclusive jet mass distribution in $e^+e^-$ events, defined with a veto on the out-of-jet radiation, at two-loop order.
In particular, we calculate the two-loop soft function, which is required to describe this distribution in the threshold
region. When combined with other ingredients using soft-collinear effective theory, this generates the complete singular distribution for jet thrust, the sum of the jet masses, at two-loop order. The result is in excellent agreement with full QCD. The integrated jet thrust distribution is found to depend in an intricate way on both the finite jet cone size, $R$, and the jet veto scale. 
The result clarifies the structure of the potentially large logarithms (both global and non-global) which arise in jet observables for the first time at this order. 
Somewhat surprisingly, we find that, in the small $R$ limit, there is a precise and simple correspondence between the non-global contribution to the integrated
jet thrust distribution and the previously calculated non-global contribution to the integrated hemisphere soft function, including
subleading terms. This suggests that the small $R$ limit may provide a useful expansion for studying other
 exclusive jet substructure observables.
}

\begin{document}
\section{Introduction} 
\label{sec:intro}
High energy scattering processes in Quantum Chromodynamics~(QCD) often involve the production of jets. 
A QCD observable of particular interest is the jet mass. Jet mass  plays an important role in many applications, such as in the identification of a boosted $W$ boson~\cite{Butterworth:2002tt,Cui:2010km} or top quark~\cite{Kaplan:2008ie,Rappoccio:2011nj}. Jet mass is also important because many event shapes studied at $e^+e^-$ colliders reduce to some measure of jet mass in particular limits. For example, thrust in the dijet limit reduces to the sum of the masses of two hemisphere jets, and heavy jet mass reduces to the larger of the two hemisphere jet masses. Thrust and heavy jet mass are two of the most well-studied $e^+e^-$ event shapes and both have been measured with extraordinary
precision~\cite{Heister:2003aj}. On the theoretical side, next-to-next-to-leading order~(NNLO) QCD calculations of these observables have been carried out for jets
in $e^+e^-$  events~\cite{GehrmannDeRidder:2005cm,GehrmannDeRidder:2007bj,GehrmannDeRidder:2007hr,Weinzierl:2008iv}. 
Unfortunately, jet physics at hadron colliders is much more difficult, both experimentally and theoretically. It is therefore still worth studying $e^+e^-$ jets, where the analytical calculations can be done, in the hope that universality properties will be found
which may ultimately make the hadron collider calculations tractable, at least in some regime.  

In the limit of small jet mass, one cannot straightforwardly make accurate predictions for the thrust or heavy jet mass using QCD perturbation theory. In this limit, the cross section is dominated by the soft and collinear interactions of massless partons which produce large logarithmic contributions of the form $\als^i \ln^j\left(m/Q\right)$ at each order in perturbation theory. Without a correct resummation of these terms to all orders in $\als$, perturbative predictions are not reliable when $m/Q \to 0$. In recent years, much progress has been made on this front and NNNLL resummations of both thrust and heavy jet mass are now available~\cite{Schwartz:2007ib,Becher:2008cf,Chien:2010kc}. These resummations led to a measurement of the strong coupling constant, $\als$, which is competitive with the world average~\cite{Abbate:2010xh}. Soft-collinear effective theory~(SCET) is the theoretical tool that facilitated many of these important developments~\cite{Bauer:2000yr,Bauer:2001yt,Beneke:2002ph,Fleming:2007qr}. Indeed, before the simplifications brought about by SCET, thrust and heavy jet mass had only been resummed at the NLL level~\cite{Catani:1992ua}. 

Although these studies have shown the power of resummation, the results for thrust and heavy jet mass are not directly applicable to the study of observables at hadron colliders. At hadron colliders, there is soft and collinear activity in the initial state that complicates the definition of jet mass~\cite{Kelley:2010qs}. In reality, experiments at hadrons colliders use inclusive jet definitions, and an explicit veto procedure to reduce the huge QCD background. Therefore, it is interesting and non-trivial to try and carry out the resummation of a QCD observable in the presence of a jet veto. In fact, some work in this direction has already been done and it is known that, in this case, a new class of so-called non-global logarithms~(NGLs), qualitatively different from the class of logarithms encountered in the resummation of thrust and heavy jet mass, appear which also need to be resummed in certain kinematical limits~\cite{Dasgupta:2001sh,Dasgupta:2002bw,Banfi:2002hw,Appleby:2002ke}. Unfortunately, the resummation of these logarithms is not straightforward (see Refs~\cite{Dasgupta:2001sh,Banfi:2002hw,Kelley:2011ng,Marchesini:2003nh,Avsar:2009yb}).
Over the last couple of years, there has been renewed theoretical interest in exclusive jet mass distributions and NGLs, both with and without SCET~\cite{Ellis:2009wj,Ellis:2010rwa,Banfi:2010pa,Kelley:2011tj,Kelley:2011ng,Hornig:2011iu,Li:2011hy,Hornig:2011tg}. 

In a previous publication, Ref.~\cite{Kelley:2011tj}, a simple observable called jet thrust was introduced. 
It was subsequently studied in Refs.~\cite{Hornig:2011tg,KhelifaKerfa:2011zu}.
Jet thrust, denoted $\tau_\omega$, is defined as the sum of the masses of the two hardest jets in an event with a veto that restricts the total out-of-jet energy to be less than $\omega$. This observable, being a single variable (once the jet size $R$ and the veto scale $\omega$ are fixed) is easy to study qualitatively with one-dimensional plots. It is also free of a subset of non-global logarithms of the form $\ln\left(m_1/m_2\right)$. In Ref.~\cite{Kelley:2011tj}, it was observed that the soft function for jet thrust seems to refactorize at small $R$. Heuristic arguments for this refactorization were given, as was a convincing numerical comparison of predictions derived from this refactorization to the singular part of the exact $\mathcal{O}(\als^2)$ results in QCD derived using the program~{\event}~\cite{Catani:1996jh,Catani:1996vz}. 
The leading non-global logarithms for jet thrust, which were ignored in Ref.~\cite{Kelley:2011tj}, were subsequently studied in Refs.~\cite{Hornig:2011tg,KhelifaKerfa:2011zu}. Understanding analytically how the refactorization conjectured in Ref.~\cite{Kelley:2011tj} 
works, as well as the structure of the subleading non-global contributions, requires the two-loop result.
The two-loop result is the subject of the present paper.

In Ref.~\cite{Kelley:2011ng}, the exact $\Ord\left(\als^2\right)$ hemisphere soft function was calculated. This facilitated
an analytical study of the hemisphere mass distribution at NLO. In particular, subleading non-global logarithms were uncovered for
the first time, as well as an intricate non-global functional dependence, in sharp contrast to the simple powers of logarithms suggested
by~\cite{Hoang:2008fs}.
Parts of this result were independently confirmed by two other groups~\cite{Hornig:2011iu,Monni:2011gb}.

In this paper, we go one step further and perform the calculation of the soft function for jet mass with a jet veto
at $\mathcal{O}(\als^2)$ using jets of cone size $R$. 
This produces an exact form for the singular part of the differential jet thrust distribution. In fact, we present the complete scale-dependent contribution to the integrated jet thrust distribution in the threshold region.
Our result is exact up to the addition of a $\tom$- and $\omega$-independent function of $R$.
This function is calculable, but does not affect the differential distribution, and is therefore not essential for our present purposes. 
Our results allow us to both test the refactorization ansatz of Ref.~\cite{Kelley:2011tj} and determine the precise form of the non-global terms which arise in the integrated jet thrust distribution. 

A number of authors have already observed that there are similarities between the leading NGLs in jet observable distributions defined using narrow anti-$k_T$ jets and the NGLs that appear in the hemisphere mass distribution~\cite{Banfi:2010pa,Kelley:2011tj,Hornig:2011iu,Hornig:2011tg}. The $R$-dependent coefficient of the leading non-global logarithm
\begin{equation}
 f(R) \ln^2\left(\frac{\omega}{\tom Q}\right)
\end{equation}
has been calculated for various algorithms, for arbitrary $R$.
However, all existing NGL studies with $R$ dependence focus on the extraction of the leading non-global logarithm, which can be
done without a complete two-loop calculation. 

In this work, we calculate the {\it complete} non-global contribution to the integrated jet thrust distribution, up to a scale-independent
function of $R$. Remarkably, in the small $R$ limit, we find that this non-global contribution to the integrated $\tau_\omega$ distribution can be expressed in terms of the non-global contribution to the integrated hemisphere soft function calculated in Ref.~\cite{Kelley:2011ng}.
The correspondence actually reproduces the entire non-global functional form, not just the non-global large logarithms.
 We also confirm that all of the singular terms in $R$ in the integrated jet thrust distribution can be reproduced by the refactorization formula of Ref.~\cite{Kelley:2011tj}, provided that one takes into account the natural $R$ dependence in the argument of the non-global function.

This paper is organized as follows. In Section \ref{sec:review}, we briefly review the factorization theorem for the $\tau_\omega$ distribution and recall the refactorization ansatz for the $\tau_\omega$ soft function proposed in Ref.~\cite{Kelley:2011tj}. In Section \ref{sec:calculation}, we summarize the calculation of the $\tau_\omega$ soft function, at times focusing only on those moments of the integrals which ultimately contribute to the scale-dependent terms in the integrated $\tau_\omega$ distribution. In Section \ref{sec:difference}, we compare the prediction of SCET to full QCD and use the difference of our results and the prediction of the integrated refactorization ansatz to define the ($\mu$-independent) non-global contribution to the integrated $\tau_\omega$ distribution. In Section \ref{sec:smallr}, we take the small $R$ limit of the non-global contribution to the integrated $\tom$ distribution. We then make precise the correspondence alluded to above between this non-global contribution and the non-global contribution to the integrated hemisphere soft function. Finally, we analyze the NGLs in detail, discuss their $R$ dependence, and present a refined refactorization formula for the $\tom$ soft function consistent with our two-loop results. In Section \ref{sec:conclusions}, we present our conclusions and discuss some interesting open problems. In Appendix \ref{app:int} we present some of the technical details of our calculation of the integrated $\tom$ distribution and in Appendix \ref{app:chi} we collect two rather lengthy analytical expressions that would have been awkward to define in the text where they first appear. A {\tt Mathematica} notebook
with these expressions is available upon request.

\section{Jet Thrust in SCET}
\label{sec:review}
Soft functions are squared matrix elements of Wilson lines integrated against some measurement operator. They are integral to QCD resummation studies and have appeared in the literature in many different contexts~\cite{Korchemsky:1985xj,Korchemsky:1993uz,Belitsky:1998tc,Aybat:2006mz}.
In Ref.~\cite{Kelley:2011tj}, an inclusive jet mass observable called jet thrust, $\tau_\omega$, was discussed, and after
making a conjecture for part of the two-loop soft function, the $\mathcal{O}(\als^2)$ prediction from SCET for the $\tau_\omega$ distribution was compared with the output of {\event}. In this section, we briefly review the definition of $\tau_\omega$ and recapitulate the main results of Ref.~\cite{Kelley:2011tj} which sets the stage for the exact results calculated in the present paper.

Jet thrust $\tau_\omega$ is defined as follows. First, for some multi-jet event at an $e^+e^-$ collider, cluster
the particles into jets using some jet algorithm with size parameter $R$. Define $\lambda$ to be the energy of the radiation
not going into the two hardest jets.\footnote{$\lambda$ can also be defined as the energy of the third hardest jet,
as it was in~\cite{Kelley:2011tj}. The definitions are equivalent definition as far as the differential $\tom$ distribution at two loops is concerned.} 
Then jet thrust is defined as the sum of the squared masses of the two hardest jets normalized to the center of mass energy if $\lambda < \omega$ and as 0 otherwise. In symbols,
\begin{equation}
\label{eq:jetomega}
\tau_\omega = \frac{M^2_1+ M^2_2}{Q^2}\,\Theta(\omega-\lambda)\,,
\end{equation}
where $M_1$ and $M_2$ are the masses of the two hardest jets and $\Theta(x)$ is the unit step function.

In the limit $\tau_\omega \ll 1$ and $\omega/Q \ll 1$, the jet thrust distribution factorizes~\cite{Kelley:2011tj}
\begin{eqnarray}
  \label{eq:factauo}
  \frac{1}{\sigma_0}\frac{\dd \sigma}{\dd \tau_\omega}
 &=& H(Q^2,\mu) \int \! \dd k_L\, 
\dd k_R\, \dd M_L^2\, \dd M_R^2\,
J(M^2_L - Q k_L,\mu) J(M^2_R-Q k_R,\mu)
\el
\times
\int^\omega_0 \! \dd \lambda \,
S_{R}(k_L,k_R,\lambda,\mu)
\delta\left( \tau_\omega - \frac{M^2_L + M^2_R}{Q^2} \right)\,,
\end{eqnarray}
up to power corrections in $\tau_\omega$ and $\omega/Q$. In Eq.~(\ref{eq:factauo}), the hard function, $H(Q^2,\mu)$, and the jet function, $J(p^2,\mu)$, are the same as the ones which appear in the factorization theorems for the thrust and heavy jet mass distributions~\cite{Ellis:2010rwa,Jouttenus:2009ns}. In principle, one could use $R$-dependent jet functions~\cite{Ellis:2009wj,Ellis:2010rwa,Jouttenus:2011wh}. However, the $R$ dependence of such jet functions shows up only in terms non-singular in the jet masses. Therefore, due to the fact that we are working in the threshold region where SCET is valid, we are free to use the simpler inclusive jet functions.

The distribution of $\tom$ depends on the precise jet definition used. In the QCD calculation, the jet definition affects the whole distribution. In the SCET calculation, it only affects the soft function. In Ref.~\cite{Kelley:2011tj}, the Cambridge/Aachen (C/A) algorithm was used. The C/A algorithm first calculates the distances between all particles $i$ and $j$
\begin{equation}
 R_{ij} = \frac{1}{2}\Big(1-\cos \theta_{i j}\Big)
\end{equation}
and then merges the two closest particles into a single particle by adding their four-momenta. The algorithm stops
when no two objects are closer than a given jet size, $R$, to one another. The Cambridge/Aachen algorithm has many appealing qualities and is infrared-safe. The corresponding $\tom$ soft function, however, appears challenging to calculate at two loops. 

Fortunately, there is a simple cone algorithm one can use instead. We define cone algorithm jet thrust as follows. First, find the thrust axis in the event. Then define the distance $R_{i\,\mathbf{n}}$ ($R_{i\,\bar{\mathbf{n}}}$) between particle $i$ in the event and the thrust axis $\mathbf{n}\,(\bar{\mathbf{n}})$ as
\begin{equation}
  \label{eq:Ri}
  R_{i\,\mathbf{n}} = \frac{1}{2}(1-\cos\theta_{i\,\mathbf{n}})\qquad\qquad R_{i\,\bar{\mathbf{n}}} = \frac{1}{2}(1-\cos\theta_{i\,\bar{\mathbf{n}}}),
\end{equation}
where $\theta_{i\,\mathbf{n}}$ ($\theta_{i\,\bar{\mathbf{n}}}$) is the angle between particle $i$ and $\mathbf{n}\,(\bar{\mathbf{n}})$. If $R_{i\,\mathbf{n}}$ ($R_{i\,\bar{\mathbf{n}}}$) is less than $R$, then cluster particle $i$ into the right (left) jet. Because we use the thrust axis, this algorithm is infrared-safe at $e^+e^-$ colliders.

The jet thrust factorization formula bears a close resemblance to the factorization formula for thrust~\cite{Schwartz:2007ib,Fleming:2007qr}. The only difference is in the soft function. In the case of thrust, the soft function is just the hemisphere one, $S_{\rm hemi}(k_L,k_R,\mu)$, and depends on two scales, $k_L$ and $k_R$. The $\tom$ soft function depends on the scales $\lambda$ and $R$ as well and it is therefore significantly more complicated. 
The soft function is defined as
\begin{eqnarray}
\label{eq:softdef}
  S_R(k_L,k_R,\lambda,\mu) &=& \frac{1}{N_c}\sum_{X_s}
\langle 0 | \bar{Y}_{\bar{n}} Y_n | X_s \rangle
\langle X_s | Y^\dagger_{n} \bar{Y}^\dagger_{\bar{n}} | 0 \rangle
\el
\times
\delta\Big(k_R-n\cdot P^R_{X_s} \Big)
\delta\Big(k_L-\bar{n}\cdot P^L_{X_s} \Big)
\delta\Big(\lambda - E_{X_s}\Big),
\end{eqnarray}
where $P^{R}_{X_s}$ $\left(P^{L}_{X_s}\right)$ is the four-momentum of the soft radiation clustered into the right (left) jet, and $E_{X_s}$ is the total energy of the out-of-jet radiation. At zeroth order, the soft function is simply a delta function
\begin{equation}
\label{eq:soft0}
S_{R}(k_L,k_R,\lambda,\mu) = \delta(k_L)\delta(k_R)\delta(\lambda),
\end{equation}
reflecting the fact that there is no soft radiation. The first non-trivial corrections come at $\Ord(\als)$ (what is traditionally
called leading-order (LO)) and were calculated in Refs.~\cite{Ellis:2010rwa} and~\cite{Kelley:2011tj}. 

Using the renormalization-group (RG) invariance of the factorization formula, one can show that the soft function must have the form
\begin{equation}
S_{R}(k_L,k_R,\lambda,\mu) = S_\mu(k_L,\mu)S_\mu(k_R,\mu)\otimes S_f(\lambda,k_L,k_R)\,, 
\label{eq:rgpred}
\end{equation}
where $\otimes$ denotes a convolution (it is a product in Laplace space).
Here $S_\mu(k,\mu)$ is the soft evolution kernel that precisely cancels the RG evolution of the jet and hard functions~\cite{Chien:2010kc,Fleming:2007qr,Hoang:2008fs}. It follows that the soft function can be written as
\begin{equation}
 S_{R}(k_L,k_R,\lambda,\mu) = S^{\in}_R(k_L,\mu)  S^{\in}_R(k_R,\mu)  S^{\out}_R(\lambda,\mu) \otimes S^f_R(\lambda,k_L,k_R)\,. \label{eq:refact}
\end{equation}
There is not actually much content in this separation until the objects involved are given precise definitions.
For example, we can take $S^\out_R(\lambda,\mu) = 1$, thus reducing Eq.~\eqref{eq:refact} to Eq.~\eqref{eq:rgpred}.
A refactorization formula like this has appeared multiple times in the literature~\cite{Ellis:2009wj,Ellis:2010rwa,Kelley:2011tj,Hornig:2011tg}. However, without a precise statement about $S^{\out}_R(\lambda,\mu)$,
Eq.~\eqref{eq:refact} actually has less content than Eq.~\eqref{eq:rgpred}.

\begin{figure}[t]
\begin{center}
$  \begin{array}{cc}
  \includegraphics[width=0.5\textwidth]{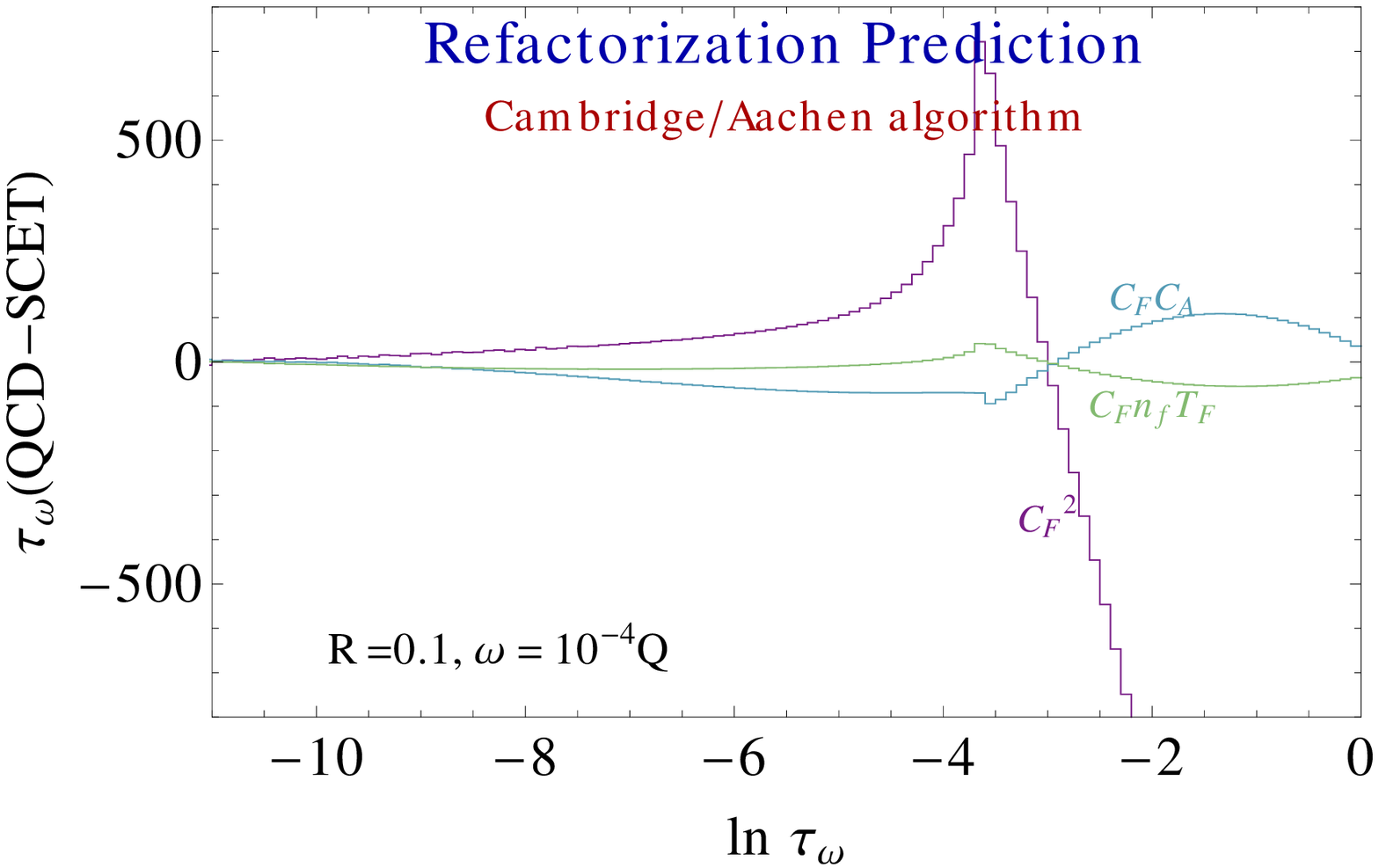}&
  \includegraphics[width=0.5\textwidth]{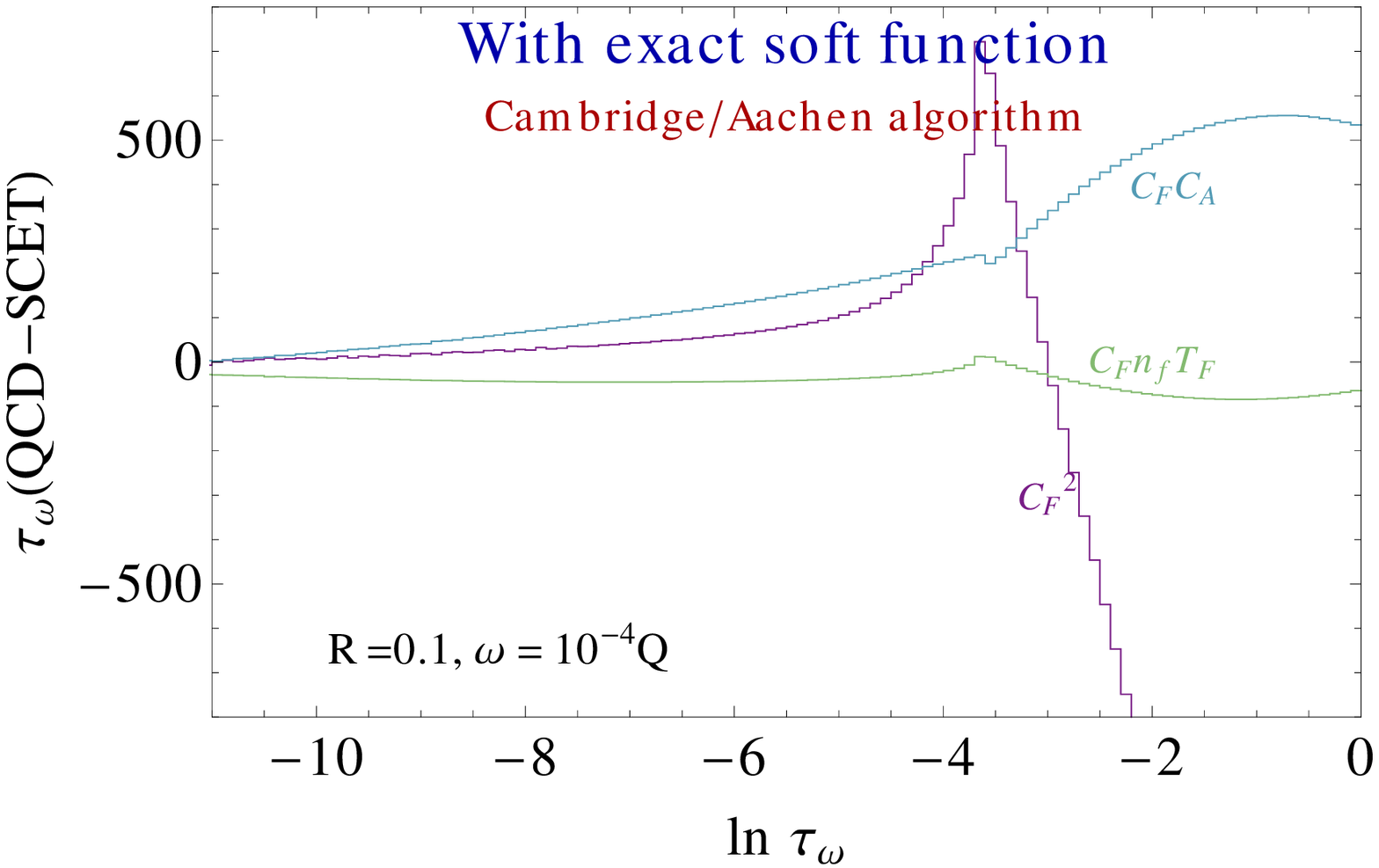}\\
\includegraphics[width=0.5\textwidth]{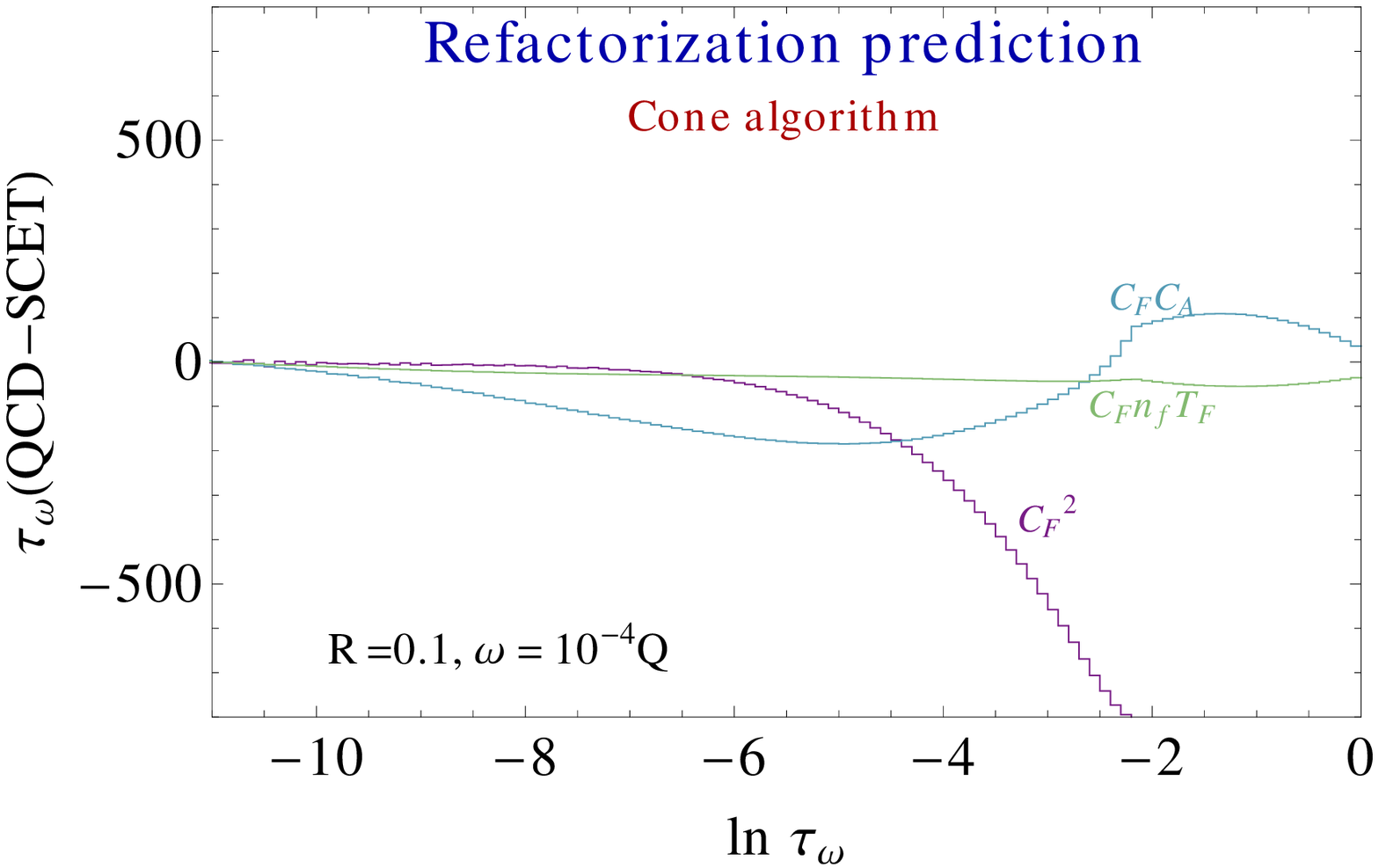}&
  \includegraphics[width=0.5\textwidth]{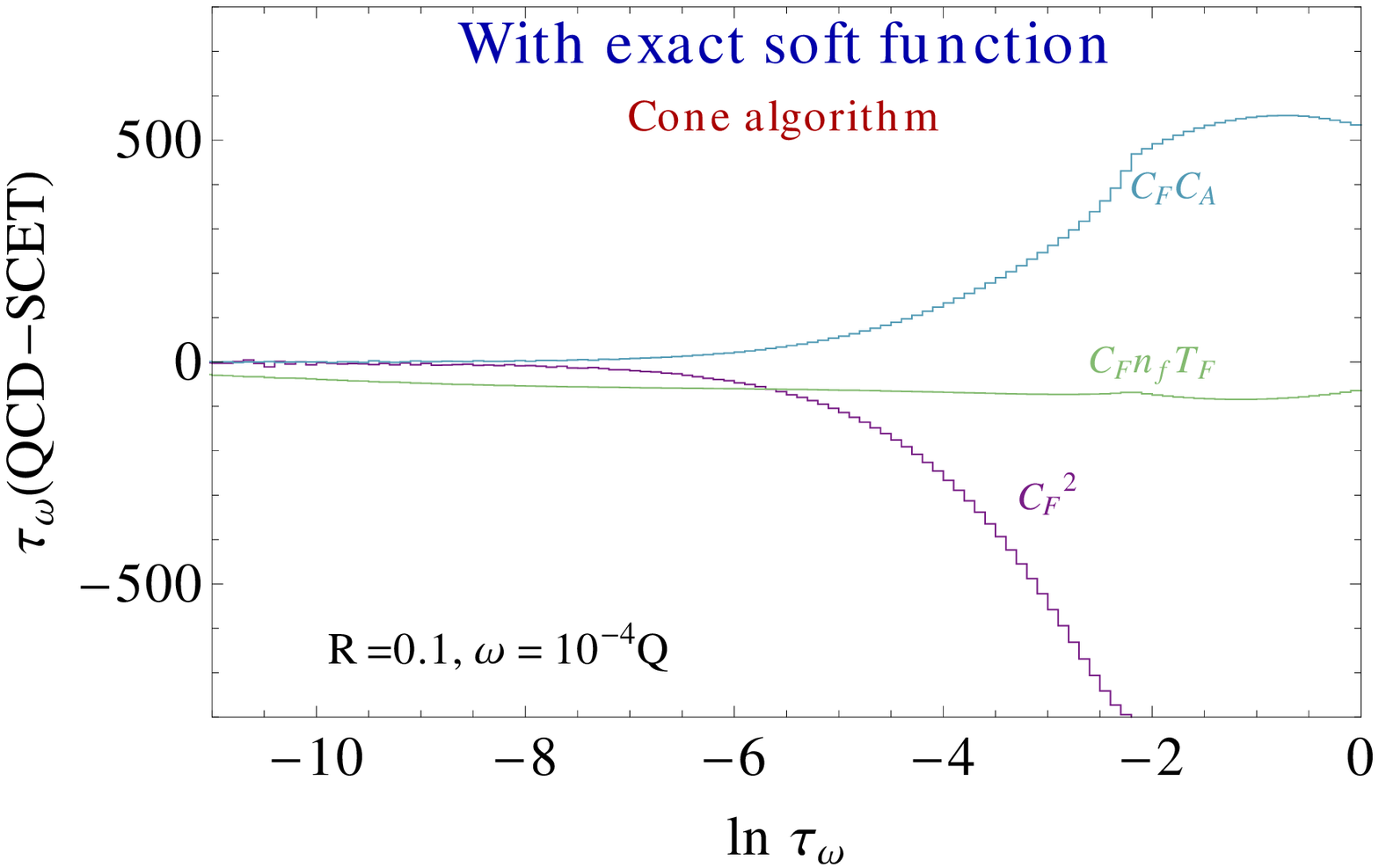}\\
  \end{array}$
\end{center}
\caption{The difference between coefficient of $\left(\frac{\alpha_s}{2\pi}\right)^2$ in $d \sigma/d \ln (\tau_\omega)$ in full QCD and in SCET. We show on the left the results obtained using the refactorization and $\gcusp$ ansatz of Ref.~\cite{Kelley:2011tj}. The results shown on the right take into account the precise $\mathcal{O}(\alpha_s^2)$ results derived in this paper. The top row uses the Cambridge/Aachen algorithm for the QCD calculation, whereas the bottom row uses a cone algorithm in QCD, in which the jet axis is taken to be the thrust direction. The top-left plot is identical to Figure 6 of Ref.~\cite{Kelley:2011tj}.
All plots have $R=0.1$ and $\omega =10^{-4} Q$.}
\label{fig:refact}
\end{figure}

In Ref.~\cite{Kelley:2011tj} it was argued that, in the small $R$ limit, $S^\in_R(k,\mu)$ and $S^\out_R(\lambda,\mu)$ should act like soft functions in their own right, with anomalous dimensions. Only the function $S^\in_R(k,\mu)$ can have a $\mu$-dependent anomalous dimension, since the soft-collinear region of phase-space lies within the jet. Thus, the anomalous dimension of $S^\in_R(k,\mu)$ has a cusp part which agrees with that of $S_\mu(k,\mu)$. In Ref.~\cite{Kelley:2011tj} it was conjectured that the regular anomalous dimensions of the in and out soft functions, at small $R$, should be
\be
\label{eq:regano}
\gamma^{\rm in}_S = \frac{1}{2}\left(\gamma_S + \gcusp\ln\left(\rr\right)\right)\qquad{\rm and}\qquad\gamma^{\rm out}_S = - \gcusp \ln\left(\rr\right)\,,
\ee
where $\gcusp$ is the usual cusp anomalous dimension and $\gamma_S$ is the regular anomalous dimension of the thrust soft function~\cite{Becher:2008cf}. Note that this ansatz has the cusp anomalous dimension generating something other than $\ln(\mu)$ in the anomalous dimension. The connection of the cusp anomalous dimension to the $R$-dependence at LO is also implicit in Refs.~\cite{Ellis:2009wj,Ellis:2010rwa}. Having $S_R^\in(k_L,\mu)$, $S_R^\in(k_R,\mu)$, and $S_R^\out(\lambda,\mu)$ together with their anomalous dimensions allows us define $S_R^f(\lambda,k_L,k_R)$ precisely: $S_R^f(\lambda,k_L,k_R)$ is everything that is left over from refactorization
with the anomalous dimensions in Eqs.~\eqref{eq:regano}.

Using the above ingredients, and setting $S_R^f(\lambda,k_L,k_R) = 1$, Ref.~\cite{Kelley:2011tj} predicted the differential $\tau_\omega$ distribution to $\mathcal{O}(\als^2)$ and explicitly compared the result to the output of {\event}. The predictions made from refactorization come very close to reproducing the output of {\event} in the regime where the factorization formula of Eq.~(\ref{eq:factauo}) is valid. In the top-left panel of Figure~\ref{fig:refact}, we reproduce Figure 6 from that paper. Figure~\ref{fig:refact} shows the differences between the Cambridge/Aachen algorithm and the cone algorithm (top and bottom rows) and the differences between the prediction of refactorization (left) and the precise $\mathcal{O}(\alpha_s^2)$ results derived in this paper (right). The plots displayed in Figure~\ref{fig:refact} take $R=0.1$, $\omega=10^{-4}Q$, and are separated by color structure.

The plots all show the difference between the differential $\tom$ distribution in full NLO QCD and the differential $\tom$ distribution in various NLO SCET-based approximations. If the singular terms in the QCD distribution are correctly reproduced by the various approximations, the plots in Figure~\ref{fig:refact} should tend to zero as $\tom \to 0$. In all four panels this vanishing is qualitatively visible. However, the fact that the bottom-right panel vanishes best is a sign that both the jet algorithm and non-global logarithms are important (but subleading) effects. For example, the leading non-global logarithm, which was set to zero in Ref.~\cite{Kelley:2011tj}, should contribute a term of the schematic form  $\ln\left(\tom\right)/\tom$  to the differential distribution. Two papers, Refs.~\cite{Hornig:2011tg} and \cite{KhelifaKerfa:2011zu}, have pointed out importance of these NGLs. Such terms, however, were beyond the scope of the analyses carried out in Ref.~\cite{Kelley:2011tj}.

To gain a better understanding of the refactorization conjecture and the NGLs, we explicitly calculate the $\tau_\omega$ soft function to $\mathcal{O}(\als^2)$ in the next section, with the understanding that our ultimate goal will be to study the scale-dependent contributions to the integrated $\tau_\omega$ distribution. As was observed in Ref.~\cite{Hornig:2011tg}, the advantage of approaching the calculation in this way is that both the $\mu$-dependent terms and NGLs of the soft contributions are completely transparent at the level of the integrated distribution. There are three color structures at $\mathcal{O}(\als^2)$: $C^2_F$, $C_F C_A$, and $C_F n_f T_F$. The $C^2_F$ terms are uniquely determined by the Abelian exponentiation theorem~\cite{Gatheral:1983cz,Frenkel:1984pz} and we therefore only present results for the $C_F C_A$ and $C_F n_f T_F$ color structures.

\section{Calculation of the $\tau_\omega$ Soft Function and Moments Thereof}
\label{sec:calculation}
The soft-function is usually expanded in $\alpha_s$ as 
\begin{equation}
S_R(k_L,k_R,\lambda,\mu) = \sum_{n=0}^{\infty} \left( \frac{\alpha_s}{4\pi} \right)^n S^{(n)}_R(k_L,k_R,\lambda,\mu).
\end{equation}
In this section we summarize the calculation of the two-loop $\tom$ soft function, $S^{(2)}_R$, at times focusing only on those moments of the soft function which will be needed later on when we discuss the jet thrust distribution and its non-global structure. We begin with a review of the one-loop soft function, establishing our notation, and then describe the NLO results.
For readers who are only interested in the interpretation of the results, this section can be skipped.

\subsection{LO (one-loop) soft-function}
\label{sec:nlosoft}
The tree-level result is trivially given as a product of delta functions, 
\begin{equation}
S^{(0)}_R(k_L,k_R,\lambda,\mu) = \delta( k_L) \delta(k_R) \delta(\lambda)\,.
\end{equation}
At first order in $\alpha_s$, the $\tau_\omega$ soft function in $d = 4 - 2\ep$ dimensions is given by
the integral\footnote{In this paper we suppress the $i\varepsilon$ prescription. 
This is appropriate since we are dealing only with real radiation.}
\begin{equation}
\label{eq:nloint}
 S_R^{(1)}(k_L,k_R,\lambda,\mu) = (4 \pi)^2 C_F \left(\frac{\mu^2 e^{\gamma_E}}{4 \pi} \right)^{\ep}
\int\! \frac{\dd^d q}{(2\pi)^d}\frac{4 i F_1(k_L,k_R,\lambda)}{q^+ q^-}\,.
\end{equation}
In Eq. (\ref{eq:nloint}), $q^+ = n\cdot q$, $q^- = \bar{n} \cdot q$, and $F_1(k_L,k_R,\lambda)$ is the 
function that implements the one-particle phase-space cuts~\cite{Cheung:2009sg}:
\bea
  \label{eq:f1}
  F_1(k_L,k_R,\lambda) &=& -2\pi i\,\delta\left(q^2\right)\Theta\left(q^0\right)
\Bigg[\Theta^q_n \delta(k_R - q^+)\delta(k_L)\delta(\lambda)
+ \Theta^q_{\bar{n}}\delta(k_L - q^-)\delta(k_R)\delta(\lambda)
\el
+ \Theta^q_{\rm out} \delta\left(\lambda-\frac{q^++q^-}{2}\right) 
\delta(k_L)\delta(k_R) \Bigg].
\eea
In Eq. (\ref{eq:f1}),
\bea
\Theta^q_n &=& \Theta \left( r q^- - q^+\right)\nn
\Theta^q_{\bar{n}} &=& \Theta \left(r q^+ - q^- \right)\nn
\Theta^q_{\rm out} &=& \Theta\left(q^- - r q^+\right)
\Theta\left(q^+ - r q^-\right)\,,
\eea
where
\begin{equation}
 r \equiv \rr \,.
\end{equation}
This last abbreviation occurs frequently throughout the rest of this paper.

Carrying out the integral in Eq. (\ref{eq:nloint}), we find that, up to and including terms of $\mathcal{O}(\als)$,
the $\tau_\omega$ soft function can be written as
\begin{equation}
S_R(k_L,k_R,\lambda,\mu)
= S^{\rm in}_R(k_L,\mu)S^{\rm in}_R(k_R,\mu)
S^{\rm out}_R(\lambda,\mu),
\end{equation}
where
\be
 S^{\in}_R(k,\mu) = \D(k) + \left(\frac{\als}{4\pi}\right)\frac{4 C_F e^{\gamma_E \e}\mu^{2\ep} r^{\ep}}{k^{1+2\ep}\ep \,\G(1-\e)}
\label{eq:sin1}
\ee
and
\bea
\label{eq:sout1}
& S^{{\rm out}}_R(\lambda,\mu) = 
\D(\lambda) + \left({\als\over 4 \pi}\right)\left( 
\frac{8 C_F e^{\gamma_E \e}\mu^{2\ep}}{(1-\ep)\ep\,\G(1-\e) (1+r)
(2\lambda)^{1+2\ep}}\right)
\left( \frac{r}{\left(1+r\right)^2}\right)^{-\ep}\\
&
\times  \Bigg\{\left(\frac{r}{1+r}\right)^\ep
\Bigg[ \ep\, _{2}F_1\left(1-\ep,1+\ep,2-\ep;\frac{1}{1+r}\right)
+ (\ep-1)\left(1+r\right)\, _2F_1\left( -\ep,\ep,1-\ep;\frac{1}{1+r}\right) \Bigg]  \nonumber \\
&- \left(\frac{1}{1+r}\right)^\ep
\Bigg[ \ep\, r\, _2F_1\left(1-\ep,1+\ep,2-\ep;\frac{r}{1+r}\right)
+
(\ep-1)(1+r)\, _2F_1\left(-\ep,\ep,1-\ep;\frac{r}{1+r}\right)\Bigg]\Bigg\}  \nonumber
\eea
are the relevant expressions calculated through $\Ord(\als)$ and to all orders in $\e$. In Eq.~(\ref{eq:sout1}), $_2F_1(a,b,c;z)$ is the Gauss hypergeometric function. Expanding the above expressions in $\e$ reproduces the results published in Ref.~\cite{Ellis:2010rwa, Kelley:2011tj}. 

The form of Eq. (\ref{eq:sout1}) suggests that the $R$ dependence of the out-of-jet soft function is not simple, even at $\mathcal{O}(\als)$. On the other hand, we note that the $\mathcal{O}(\als)$ $R$ dependence of $S^{{\rm in}}_R(k,\mu)$ is just $r^{\ep}$; it factorizes from the rest of the $\Ord(\als)$ expression. It turns out that the full answer is simply $r^\e$ times the result obtained several years ago~\cite{Schwartz:2007ib,Fleming:2007qr} for the LO hemisphere soft function. As will be shown below, a similar phenomenon occurs at NLO for the part of the in-jet soft function that dominates in the small $R$ limit. Stated more precisely, at $\mathcal{O}(\als^n)$ the contribution with $n$ soft partons clustered into the same jet is equal to the analogous contribution to the hemisphere soft function multiplied by a factor of $r^{n\ep}$. This simple relation has its origin in the fact that the integral representation of this part of the $\mathcal{O}(\als^n)$ $\tom$ soft function has nice transformation properties under the rescaling
\begin{equation}
\label{eq:redefintion}
n_\mu = \sqrt{r} n'_\mu\qquad\qquad \bar{n}_\mu = \frac{1}{\sqrt{r}}\bar{n}'_\mu\,.
\end{equation}
This observation will be discussed in more detail in Section \ref{sec:smallr}.

\subsection{NLO (two-loop) soft function}
\label{sec:nnlosoft}

In this section we present the calculation of the NLO $\tau_\omega$ soft function, focusing at times on those moments of the soft function which we will need in subsequent sections. A moment's thought reveals that the squared matrix elements written down for the hemisphere soft function in Ref.~\cite{Kelley:2011ng} can be used for the $\tau_\omega$ soft function as well. The two-particle phase-space cuts, however, are significantly more complicated. As in the hemisphere case, it is convenient to split the calculation up according to how many partons cross the final state cut.  

\subsubsection{One-parton contributions}
\label{subsubsec}

There are two different classes of contributions which have a single gluon crossing the final state cut, the real-virtual contributions and the contributions proportional to the LO $\tau_\omega$ soft function, derived by expanding the charge renormalization constant to $\Ord(\als)$. In what follows, $S^{\in,\,(1)}_R(k,\mu)$ and $S^{{\rm out},\,(1)}_R(\lambda,\mu)$ are simply the $\Ord(\als)$ terms in Eqs. (\ref{eq:sin1}) and (\ref{eq:sout1}) respectively.

The real-virtual interference terms can be derived by judiciously combining the analogous result derived in Ref.~\cite{Kelley:2011ng} for the hemisphere soft function and the all-orders-in-$\e$ LO results collected in Section \ref{sec:nlosoft}. We find that the result can be written as
\begin{eqnarray}
S^{R-V}_R(k_L,k_R,\lambda,\mu) &=& S_{C_A}^{V}(k_L,k_R)\, r^{2\e}\,\delta(\lambda) + C_A\,S^{{\rm out},\,(1)}_R(\lambda,\mu)\bigg|_{\ep \to 2\ep}\delta(k_L)\delta(k_R)
\nonumber \\
&&\times{\pi \G(2+\e)\G(1-2\e)\cot(\pi \e)\G(-\e)^2\over \G(1-\e)\G(-2\e)(1+\e)}\,, 
\end{eqnarray}
where $S_{C_A}^V(k_L,k_R)$ is defined in Eqs. (20) and (21) of Ref.~\cite{Kelley:2011ng}. It is worth pointing out that the in-jet terms, $S_{C_A}^V(k_L,k_R)\, r^{2\e}\,\delta(\lambda)$, are simply derived by appropriately rescaling the real-virtual hemisphere integrals. For the sake of completeness, we recapitulate the result for $S_{C_A}^V(k_L,k_R)$ derived in  Ref.~\cite{Kelley:2011ng}:
\begin{eqnarray}
 S^{\rm V}_{C_A}(k_L, k_R) =   C_F \,C_A\, \mu^{4\e} \left(\frac{\delta(k_R)}{k_L^{1+4\e}} + \frac{\delta(k_L)}{k_R^{1+4\e}}\right)
 \left(-\frac{4}{\epsilon^3}
 +\frac{2\pi^2}{\epsilon} 
 +\frac{32 \zeta_3}{3}
 -\e \frac{\pi ^4}{30} + \Ord\left(\e^2\right)\right)\,.\qquad
\end{eqnarray}

Given the results of Section \ref{sec:nlosoft}, it is trivial to write down the charge renormalization contributions to the NLO $\tau_\omega$ soft function:
\bea
S^{\rm Ren}_R(k_L,k_R,\lambda,\mu)
&=&
 -
\frac{\beta_0}{\ep}
\Bigg[\left(S^{{\rm in},\,(1)}_R(k_L,\mu)
\delta(k_R)+ S^{{\rm in},\,(1)}_R(k_R,\mu)
\delta(k_L)\right)\D(\lambda)
\nonumber \\
&&+ S^{{\rm out},\,(1)}_R(\lambda,\mu)\delta(k_L)\delta(k_R)\Bigg],
\eea
where $\beta_0=\frac{11}{3}C_A - \frac{4}{3} n_f T_F$ is the leading order $\beta$-function. 
\subsubsection{Two-parton contributions}
We now turn to the contributions with two partons crossing the final state cut. The sum of all such contributions to the $\tau_\omega$ soft function has the form:
\cmb{-1 cm}{0 cm}
\bea
    S^{\rm Real}_R(k_L,k_R,\lambda,\mu) &=& (4\pi)^4 \left(\frac{\mu^2 e^{\gamma_E}}{4\pi}\right)^{2\ep}
\int\! \frac{\dd^d q}{(2\pi)^d}
\frac{\dd^d k}{(2\pi)^d}\left(\frac{1}{2!}I_{C_A} +I_{n_f}\right)F_2(k_L,k_R,\lambda)\,,
\label{eq:init2}
\eea
\cme
where $\frac{1}{2!}I_{C_A}$ and $I_{n_f}$ are the integrands for the $C_F C_A$ and $C_F n_f T_F$ color structures,
for which explicit formulas can be found in  Eqs. (16) and (27) of Ref.~\cite{Kelley:2011ng}.
The statistical factor of $1/2!$ in front of $I_{C_A}$ has its origin in the fact that, for the $C_F C_A$ color structure, the two-parton contributions have two indistinguishable gluons in the final state.
 As mentioned above, the two-parton phase-space cuts are much more complicated than the one-parton ones:
\bea
\label{eq:f2}
&&F_2(k_L,k_R,\lambda) =(-2\pi i)^2\delta\left(k^2\right) \Theta\left(k^0\right)\delta\left(q^2\right)\Theta\left(q^0\right)
\el\times\Bigg[
\Theta^q_n \Theta^k_n
 \delta(k_R - k^+ - q^+)\delta(k_L)\delta(\lambda)+\Theta^q_\bn\Theta^k_\bn \delta(k_L - k^- - q^-)\delta(k_R)\delta(\lambda)
\el
+ \Theta^k_n \Theta^q_\bn
\delta(k_R - k^+)\delta(k_L - q^-)\delta(\lambda)+
\Theta^q_n\Theta^k_\bn \delta(k_L - k^-)\delta(k_R - q^+)\delta(\lambda)
\el
+ \Theta^k_n \Theta^q_{\rm out} \delta(k_R - k^+)\delta\left(\lambda - \frac{q^++q^-}{2}\right) \delta(k_L)+ \Theta^q_n \Theta^k_{\rm out} \delta(k_R - q^+)\delta\left(\lambda - \frac{k^++k^-}{2}\right) \delta(k_L)
 \el
+ \Theta^k_\bn \Theta^q_{\rm out} \delta(k_L - k^-)\delta\left(\lambda - \frac{q^++q^-}{2}\right) \delta(k_R)+
 \Theta^q_\bn \Theta^k_{\rm out} \delta(k_L - q^-)\delta\left(\lambda - \frac{k^++k^-}{2}\right) \delta(k_R)
 \el
 +
 \Theta^k_{\rm out} \Theta^q_{\rm out} \delta
 \left(\lambda - \frac{q^+ + k^+ + q^- + k^-}{2}\right)\delta(k_L)\delta(k_R)
 \Bigg].
\eea
There are four independent cases to consider:
\begin{enumerate}
 \item  The same-side in-in contributions, where both partons are in the same jet (the second line of Eq.~\eqref{eq:f2}).
\item The opposite-side in-in contributions, where one parton is in the $\mathbf{n}$ jet and the other is in the $\overline{\mathbf{n}}$ jet
(the third line of Eq.~\eqref{eq:f2}).
\item The in-out contributions, where one parton is in a jet and the other is outside of both jets (the fourth and fifth lines of Eq.~\eqref{eq:f2}).
\item The out-out contribution, where both partons are outside of all jets (the last line of Eq.~\eqref{eq:f2}).
\end{enumerate}
%
We consider each class of contributions in turn.

Na\"{i}vely, one might expect the calculation of the same-side in-in contributions to be challenging since the evaluation of the analogous hemisphere integrals was by far the most technically demanding part of the calculation described in Ref.~\cite{Kelley:2011ng}. Fortunately, it turns out that we can recycle the same-side in-in contributions to the hemisphere soft function and obtain the desired result for free. The result of interest can be derived by appropriately rescaling the analogous hemisphere integrals. For arbitrary $R$ we find
\bea
\label{eqs1}
&&S^{r_1}_R(k_L,k_R,\lambda,\mu)=
\left({\D(k_R)\over k_L^{1+4\e}} + {\D(k_L)\over k_R^{1+4\e}}\right)\mu^{4\e} r^{2\e}\D(\lambda)
\Bigg\{
C_F C_A\Bigg[
\frac{4}{\e^3}+\frac{22}{3\e^2} +\left(\frac{134}{9}\right.
\el
\left.-\frac{4\pi^2}{3}
\right)\frac{1}{\e} + \frac{772}{27}+\frac{11\pi^2}{9}-{116 \zeta_3\over 3}
+ \ep\left( \frac{4784}{81} + \frac{67\pi^2}{27} - \frac{137\pi^4}{90}
+ \frac{484\zeta_3}{9} \right) + \Ord\left(\e^2\right)\Bigg]
\el
+ C_F n_f T_F
\Bigg[ -\frac{8}{3\e^2} - \frac{40}{9\e} - \frac{152}{27} - \frac{4\pi^2}{9} - \e \left(\frac{952}{81} + \frac{20\pi^2}{27} + \frac{176\zeta_3}{9}\right) + \Ord\left(\e^2\right)\Bigg]\Bigg\}\,,
\eea
where we have made use of Eqs. (18) and (29) in Ref.~\cite{Kelley:2011ng}. 

For the opposite-side in-in contributions, the result can be expressed as
\bea
S^{r_2}_R(k_L,k_R,\lambda,\mu)
&=&
\frac{\mu^{4\e}}{(k_L k_R)^{1+2\ep}}\,\delta(\lambda)
\el
\times
\left[
C_F C_A\,\mbox{\Large $f$}_{C_A}\left(\frac{k_L}{k_R},r\right)
+ 
C_F n_f T_F\,\mbox{\Large $f$}_{n_f} \left(\frac{k_L}{k_R},r\right)
\right],
\eea
where $\mbox{\large $f$}_{C_A}\left(z,r\right)$ and $\mbox{\large $f$}_{n_f}\left(z,r\right)$ are functions with $\ep$ expansions that begin at $\Ord\left(\e^0\right)$. In other words,
\bea
&&\mbox{\Large $f$}_{C_A}\left(z,r\right) = \mbox{\Large $f$}_{C_A}^{(0)}\left(z,r\right) + \mbox{\Large $f$}_{C_A}^{(1)}\left(z,r\right) \e + \Ord\left(\e^2\right)
\el
\mbox{\Large $f$}_{n_f}\left(z,r\right) = \mbox{\Large $f$}_{n_f}^{(0)}\left(z,r\right) + \mbox{\Large $f$}_{n_f}^{(1)}\left(z,r\right) \e + \Ord\left(\e^2\right)\,.
\label{eq:fmoms}
\eea
It is worth pointing out that, in the limit $R\to 0$, these functions are suppressed relative to the other
contributions discussed in this section. In order to obtain exact results for the scale-dependent contributions to the integrated $\tau_\omega$ distribution, we need several moments of the functions defined implicitly above:
\bea
\mbox{\Large $f$}_{C_A}^{(0)}\left(0,r\right) &=&  16 {\rm Li}_2\left(r^2\right)\,,\nn
\mbox{\Large $f$}_{n_f}^{(0)}\left(0,r\right) &=& 0\,,\nn
\mbox{\Large $f$}_{C_A}^{(1)}\left(0,r\right) &=& -16 {\rm Li}_3\left(1-r^2\right)+16 {\rm Li}_3\left(\frac{r^2}{r^2-1}\right)+32 {\rm Li}_2\left(r^2\right) \ln \left(\frac{r}{1-r^2}\right)
\el
-\frac{8}{3} \ln^3\left(1-r^2\right)-16 \ln (r) \ln^2\left(1-r^2\right)+\frac{8}{3} \pi ^2 \ln \left(1-r^2\right)+16 \zeta_3\,,\nn
\mbox{\Large $f$}_{n_f}^{(1)}\left(0,r\right) &=& 0\,,\nonumber
\eea
\bea
&&\int_0^1  {\dd z\over z}\left({2\mbox{\Large $f$}_{C_A}^{(0)}\left({z \over 2-z},r\right)\over 2-z} - \mbox{\Large $f$}_{C_A}^{(0)}\left(0,r\right)\right) = -\frac{88 {\rm Li}_2\left(r^2\right)}{3}+8 {\rm Li}_3\left(r^2\right)
\el
-8 {\rm Li}_2\left(r^2\right) \ln (r)+16 \ln (2) {\rm Li}_2\left(r^2\right)+\frac{8
   r^2}{3 \left(r^2-1\right)}-\frac{16 r^2 \ln (r)}{3 \left(r^2-1\right)^2}-\frac{88}{3} \ln (r) \ln \left(1-r^2\right)\,,\nonumber
\eea
and
\bea
&&\int_0^1  {\dd z\over z}\left({2\mbox{\Large $f$}_{n_f}^{(0)}\left({z \over 2-z},r\right)\over 2-z} - \mbox{\Large $f$}_{n_f}^{(0)}\left(0,r\right)\right) = \frac{32 {\rm Li}_2\left(r^2\right)}{3}-\frac{16 r^2}{3 \left(r^2-1\right)}\nn
&&\qquad\qquad\qquad\qquad\qquad\qquad\qquad\qquad\qquad+\frac{32 r^2 \ln (r)}{3 \left(r^2-1\right)^2}+\frac{32}{3} \ln (r) \ln
   \left(1-r^2\right)\,.
\label{momentsinin}
\eea

The calculation of the in-out contributions proceeds similarly. However, in this case, the results are significantly more complicated. As we shall see, these contributions lead to terms in the integrated $\tau_\omega$ distribution that depend on ${\tau_\omega Q \over 2 \omega}$ in a highly non-trivially way. The result can be written as
\begin{eqnarray}
\label{eq:r4}
S^{r_3}_R(k_L,k_R,\lambda,\mu) &=& 
\frac{\mu^{4\ep}}{(2\lambda)^{1+2\ep}}\Bigg\{\frac{\delta(k_R)}{k_L^{1+2\ep}}
\left[ C_F C_A\, \mbox{\Large $g$}_{C_A}\left(\frac{k_L}{2\lambda},r\right)\right.
\el
\left.+
C_F n_f T_F\, \mbox{\Large $g$}_{n_f}\left(\frac{k_L}{2\lambda},r\right)\right]+ (k_L \leftrightarrow k_R)\Bigg\}\,,
\end{eqnarray}
where $\mbox{\large $g$}_{C_A}\left(z,r\right)$ and $\mbox{\large $g$}_{n_f}\left(z,r\right)$ are functions with $\ep$ expansions that begin at $\Ord\left(\e^0\right)$. In other words,
\bea
&&\mbox{\Large $g$}_{C_A}\left(z,r\right) = \mbox{\Large $g$}_{C_A}^{(0)}\left(z,r\right) + \mbox{\Large $g$}_{C_A}^{(1)}\left(z,r\right) \e + \Ord\left(\e^2\right)
\el
\mbox{\Large $g$}_{n_f}\left(z,r\right) = \mbox{\Large $g$}_{n_f}^{(0)}\left(z,r\right) + \mbox{\Large $g$}_{n_f}^{(1)}\left(z,r\right) \e + \Ord\left(\e^2\right)\,.
\label{eq:gmoms}
\eea
In order to obtain exact results for the scale-dependent contributions to the integrated $\tau_\omega$ distribution, we need several moments of the functions defined implicitly above:
\bea
\mbox{\Large $g$}_{C_A}^{(0)}\left(0,r\right) &=&  \frac{16 \pi ^2}{3}-64 {\rm Li}_2\left(r\right)-64 {\rm Li}_2\left(-r\right)\,,\nn
\mbox{\Large $g$}_{n_f}^{(0)}\left(0,r\right) &=& 0\,,\nn
\mbox{\Large $g$}_{C_A}^{(1)}\left(0,r\right) &=& -16 {\rm Li}_3\left(r^2\right)+128 {\rm Li}_3(1-r)+128 {\rm Li}_3\left(\frac{r}{r+1}\right)-128 {\rm Li}_2(-r) \ln (r+1)
\el
+64 {\rm Li}_2(-r) \ln (r)+128
   {\rm Li}_2(r) \ln (1-r)-\frac{64}{3} \ln ^3(r+1)+64 \ln ^2(1-r) \ln (r)
\el
-\frac{64}{3} \pi ^2 \ln (1-r)-96 \zeta_3\,,\nn
\mbox{\Large $g$}_{n_f}^{(1)}\left(0,r\right) &=& 0\,,\nonumber
\eea
\bea
&&\int_0^1{\dd z\over z}\left(\mbox{\Large $g$}_{C_A}^{(0)}\left(z,r\right) + \mbox{\Large $g$}_{C_A}^{(0)}\left(1/z,r\right) - 2 \mbox{\Large $g$}_{C_A}^{(0)}\left(0,r\right)\right) = \frac{352 {\rm Li}_2\left(r^2\right)}{3}-16
   {\rm Li}_3\left(r^2\right)
\el
-64 {\rm Li}_3(r)+64 {\rm Li}_3\left(\frac{1}{r+1}\right)+64 {\rm Li}_3\left(\frac{r}{r+1}\right)+64 {\rm Li}_2(-r) \ln (r)+64{\rm Li}_2(r) \ln (r)
\el
-\frac{16 \left(r^2+1\right)}{3\left(r^2-1\right)}+\frac{64 r^2 \ln (r)}{3\left(r^2-1\right)^2}-\frac{64}{3} \ln ^3(r+1)+32 \ln (r) \ln^2(r+1)+\frac{352}{3} \ln (r) \ln (r+1)
\el
\qquad\qquad\qquad\qquad\qquad\qquad\qquad+\frac{32}{3} \pi ^2 \ln (r+1)+\frac{352}{3} \ln (1-r) \ln (r)-32 \zeta_3-\frac{176 \pi^2}{9}\,,
\el
\int_0^1{\dd z\over z}\left(\mbox{\Large $g$}_{n_f}^{(0)}\left(z,r\right) + \mbox{\Large $g$}_{n_f}^{(0)}\left(1/z,r\right) - 2 \mbox{\Large $g$}_{n_f}^{(0)}\left(0,r\right)\right) = -\frac{128 {\rm Li}_2\left(r^2\right)}{3}-\frac{128 r^2 \ln (r)}{3
   \left(r^2-1\right)^2}
\el
\qquad\qquad\qquad\qquad\qquad\qquad\qquad\qquad\qquad+ \frac{64 \pi^2}{9} + \frac{32 \left(r^2+1\right)}{3 \left(r^2-1\right)}-\frac{128}{3} \ln (r) \ln \left(1-r^2\right)\,,
\el
\int_1^{x} {\dd z_2 \over z_2}\int_0^1  {\dd z_1\over z_1}\left(\mbox{\Large $g$}_{C_A}^{(0)}\left({z_2 \over z_1},r\right) - \mbox{\Large $g$}_{C_A}^{(0)}\left(0,r\right)\right) = \mbox{\Large $\chi$}_{C_A}\left(x,r\right) - \mbox{\Large $\chi$}_{C_A}\left(1,r\right)\,,\nonumber
\eea
and
\bea
&&\int_1^{x} {\dd z_2 \over z_2}\int_0^1  {\dd z_1\over z_1}\left(\mbox{\Large $g$}_{n_f}^{(0)}\left({z_2\over z_1},r\right) - \mbox{\Large $g$}_{n_f}^{(0)}\left(0,r\right)\right) = \mbox{\Large $\chi$}_{n_f}\left(x,r\right) - \mbox{\Large $\chi$}_{n_f}\left(1,r\right)\,.
\label{momentsinout}
\eea
The functions $\mbox{\large $\chi$}_{C_A}\left(x,r\right)$ and $\mbox{\large $\chi$}_{n_f}\left(x,r\right)$ are given in Appendix \ref{sec:anacumulant} in terms of an appropriate set of one- and two-dimensional harmonic polylogarithms.\footnote{For the reader less familiar with harmonic polylogarithms, we recommend reading Ref. \cite{Remiddi:1999ew} and Appendix A of Ref. \cite{Gehrmann:2000zt}.}
 
Finally, for the out-out contribution, the result is
\bea
&&S^{r_4}_R(k_L,k_R,\lambda,\mu) =
\frac{\mu^{4\ep}}{(2\lambda)^{1+4\ep}}
\delta(k_L)\delta(k_R)
\Bigg\{
C_F C_A
\Bigg[
-\frac{32 \ln (r)}{\ep^2}
+
\frac{1}{\ep}
\Bigg(
16\pi^2 
\el
- \frac{176\ln (r)}{3}+ 32\ln^2 (r) + 64 \lib (-r) - 64\lib (r)
\Bigg)+ 64 {\rm Li}_3\left(\frac{r^2}{r^2-1}\right)-\frac{704
   {\rm Li}_2(r)}{3}
\el
+256 {\rm Li}_3(1-r)+128 {\rm Li}_3(r)+128
   {\rm Li}_3\left(\frac{1}{r+1}\right)-128
   {\rm Li}_3\left(\frac{r}{r+1}\right)-128 {\rm Li}_2(-r) \ln (1-r)
\el
+128
   {\rm Li}_2(r) \ln (1-r)-64 {\rm Li}_2(-r) \ln (r)+128
   {\rm Li}_2(-r) \ln (r+1)+64 {\rm Li}_2(r) \ln (r)
\el
-128 {\rm Li}_2(r)
   \ln (r+1)-\frac{16 \left(r^2+1\right)}{3-3 r^2}-\frac{64 r^2 \ln
   (r)}{3 \left(r^2-1\right)^2}-\frac{32}{3} \ln ^3(1-r)-\frac{64 \ln
   ^3(r)}{3}
\el
-\frac{32}{3} \ln ^3(r+1)+128 \ln (r) \ln ^2(1-r)-32 \ln
   (r+1) \ln ^2(1-r)-32 \ln ^2(r+1) \ln (1-r)
\el
+\frac{176 \ln^2(r)}{3}+64 \ln (r) \ln ^2(r+1)-\frac{352}{3} \ln (r) \ln
   (1-r)-\frac{128}{3} \pi ^2 \ln (1-r)+\frac{32}{3} \pi ^2 \ln
   (r)
\el
-\frac{1072 \ln (r)}{9}-\frac{352}{3} \ln (r) \ln
   (r+1)+\frac{64}{3} \pi ^2 \ln (r+1)-128 \zeta_3+\frac{352 \pi ^2}{9} + \Ord(\e)\Bigg]
\el
+
C_F n_f T_F
\Bigg[
\frac{64\ln (r)}{3\ep}
+\frac{256 \mathrm{Li}_2(r)}{3}+\frac{32 (1 + r^2)}{3
   \left(1-r^2\right)}
+\frac{128
   r^2 \ln (r)}{3 \left(1-r^2\right)^2}-\frac{64 \ln^2(r)}{3}
+\frac{320 \ln (r)}{9}
\el
+\frac{128}{3} \ln (r) \ln
   \left(\frac{1-r}{r+1}\right)
+\frac{256}{3} \ln (r) \ln
   (r+1)-\frac{128 \pi ^2}{9} + \Ord(\e)\Bigg]
\Bigg\}.
\label{eq:outout}
\eea
The calculation of this contribution is actually quite involved; in the integrals that lead to Eq. (\ref{eq:outout}), there is explicit $R$ dependence (which does not factorize from the rest of the expression) and one has to use sector decomposition~\cite{Binoth:2000ps} in a non-trivial way to deal with the phase-space singularities. As we shall see later on, integrating the out-out contribution to the $\tau_\omega$ soft function leads to logarithms of $\frac{\mu}{2\omega}$. We expanded the non-trivial part of the right-hand side of Eq. (\ref{eq:outout}) to $\mathcal{O}\left(\e^0\right)$ because it turns out that this completely determines the logarithms of $\frac{\mu}{2\omega}$ which appear in the integrated distribution; the general arguments of~\cite{Kelley:2011tj} forbid terms that go like $\ln^3\left(\frac{\mu}{2\omega}\right)$ or $\ln^4\left(\frac{\mu}{2\omega}\right)$. 

Let us now briefly summarize what we have accomplished. In this section, various contributions to the NLO $\tau_\omega$ soft function were calculated. The full result, $S^{(2)}_{R} (k_L, k_R, \lambda, \mu)$, is naturally written as a sum of six contributions:
\bea
\label{eq:contlist}
S^{(2)}_{R} (k_L, k_R, \lambda, \mu)  &=& S^{R-V}_R(k_L,k_R,\lambda,\mu) + S^{\rm Ren}_R(k_L,k_R,\lambda,\mu) + S^{\rm r_1}_R(k_L,k_R,\lambda,\mu)  \nonumber \\
&& + S^{\rm r_2}_R(k_L,k_R,\lambda,\mu)+ S^{\rm r_3}_R(k_L,k_R,\lambda,\mu) + S^{\rm r_4}_R(k_L,k_R,\lambda,\mu).
\eea
The three terms in the first line of Eq. \eqref{eq:contlist} were easily calculated using known results (computed in Refs. \cite{Kelley:2011tj} and \cite{Kelley:2011ng}). The three terms in the last line were more involved. Instead of computing $S^{\rm r_2}_R(k_L,k_R,\lambda,\mu)$ and $S^{\rm r_3}_R(k_L,k_R,\lambda,\mu)$ directly, we instead computed the moments of those contributions relevant to the integrated jet mass distribution. We also chose not to compute the $\Ord\left(\e\right)$ coefficient of $S^{\rm r_4}_R(k_L,k_R,\lambda,\mu)$, which
only affects the constant part of the integrated distribution. The results given in this section 
determine the unknown scale-dependent terms in the integrated $\tau_\omega$ distribution in the threshold limit at NLO. 
In the next section, we examine and check these results.

\section{The Jet Thrust Distribution at Two-Loops}
\label{sec:difference}
In the previous section, the parts of the two-loop $\tom$ soft function relevant for the differential jet thrust distribution were calculated in dimensional regularization. We have chosen not to compute the parts of the soft function that contribute only to the zero bin ($\tom = 0$) of the differential jet thrust distribution. For the hemisphere soft function, these terms were calculated (see Ref.~\cite{Kelley:2011ng}) because they are important for precision $\alpha_s$ measurements. However, in this paper, we are primarily interested in understanding the structure of the non-global contributions to the integrated $\tom$ distribution and in testing the refactorization conjecture of Ref.~\cite{Kelley:2011tj}. We therefore followed the approach of Ref.~\cite{Hornig:2011tg} in Section \ref{sec:calculation} and calculated only those contributions to the soft function which depend on ratios of dimensionful scales.

The triply differential jet mass cross section is a complicated singular distribution which does not have a simple expression
in momentum space. Although the two-loop integrated jet thrust distribution is not simple either for finite $R$, it is quite a bit easier to work with in practice due to the fact it is free of distributions. In terms of the various contributions to the $\tom$ soft function studied in Section~\ref{sec:calculation}, we can write the integrated jet mass distribution as
\bea
\label{eq:cumulative}
K_R^{(2)}(\tau_\omega,\omega,\mu) &=& \int^{\tau_\omega}_0 \! \dd \tau_\omega^\prime \ \int^\omega_0 \dd \lambda
\int^\infty_0 \dd k_L \dd k_R S^{(2)}_R(k_L,k_R,\lambda,\mu) \delta\left(\tau_\omega^\prime-{k_L + k_R\over Q}\right)\nn
&=&\int^{\tau_\omega}_0 \! \dd \tau_\omega^\prime \ \int^\omega_0 \dd \lambda
\int^\infty_0 \dd k_L \dd k_R \Big[S^{R-V}_R(k_L,k_R,\lambda,\mu)  \nn
&&+ S^{\rm Ren}_R(k_L,k_R,\lambda,\mu) + S^{\rm r_1}_R(k_L,k_R,\lambda,\mu)+ S^{\rm r_2}_R(k_L,k_R,\lambda,\mu)\nn
&& + S^{\rm r_3}_R(k_L,k_R,\lambda,\mu) + S^{\rm r_4}_R(k_L,k_R,\lambda,\mu)\Big]\delta\left(\tau_\omega^\prime-{k_L + k_R\over Q}\right)\,,
\eea
where the superscript $(2)$ reminds us that we are calculating the integrated distribution at NLO. The scale-dependent parts of the six terms in this expression are given explicitly in Appendix~\ref{app:int}. When these soft contributions to the  integrated $\tom$ distribution are appropriately combined with the contributions coming from the hard and jet functions, the result is a concrete prediction for the integrated $\tom$ distribution in the threshold region. This prediction should be valid for arbitrary $R$, up to power corrections in $\tom$ and $\omega/Q$.

\subsection{Comparison to full QCD}
We would like to compare our analytic results for the threshold limit to full QCD, for which the distribution is only known numerically.
To do so, we first differentiate $K_R^{(2)}(\tau_\omega, \omega, \mu)$ with respect to $\tom$ to determine the NLO part of the differential distribution. This result is then appropriately combined with the LO contributions coming from the hard and jet functions, as prescribed by the factorization formula, Eq.~(\ref{eq:factauo}). The resulting expression is the threshold prediction for $\frac{\tau_\omega}{\sigma_0}\frac{\dd \sigma}{\dd \tau_\omega}$.  This prediction can then be expanded to $\Ord\left(\als^2\right)$ and numerically compared to QCD using the program {\event}.  The details of the $\als$ expansion can be found in {\it e.g.} Ref.~\cite{Kelley:2011tj}.

\begin{figure}[t]
\begin{center}
$  \begin{array}{cc}
  \includegraphics[width=0.5\textwidth]{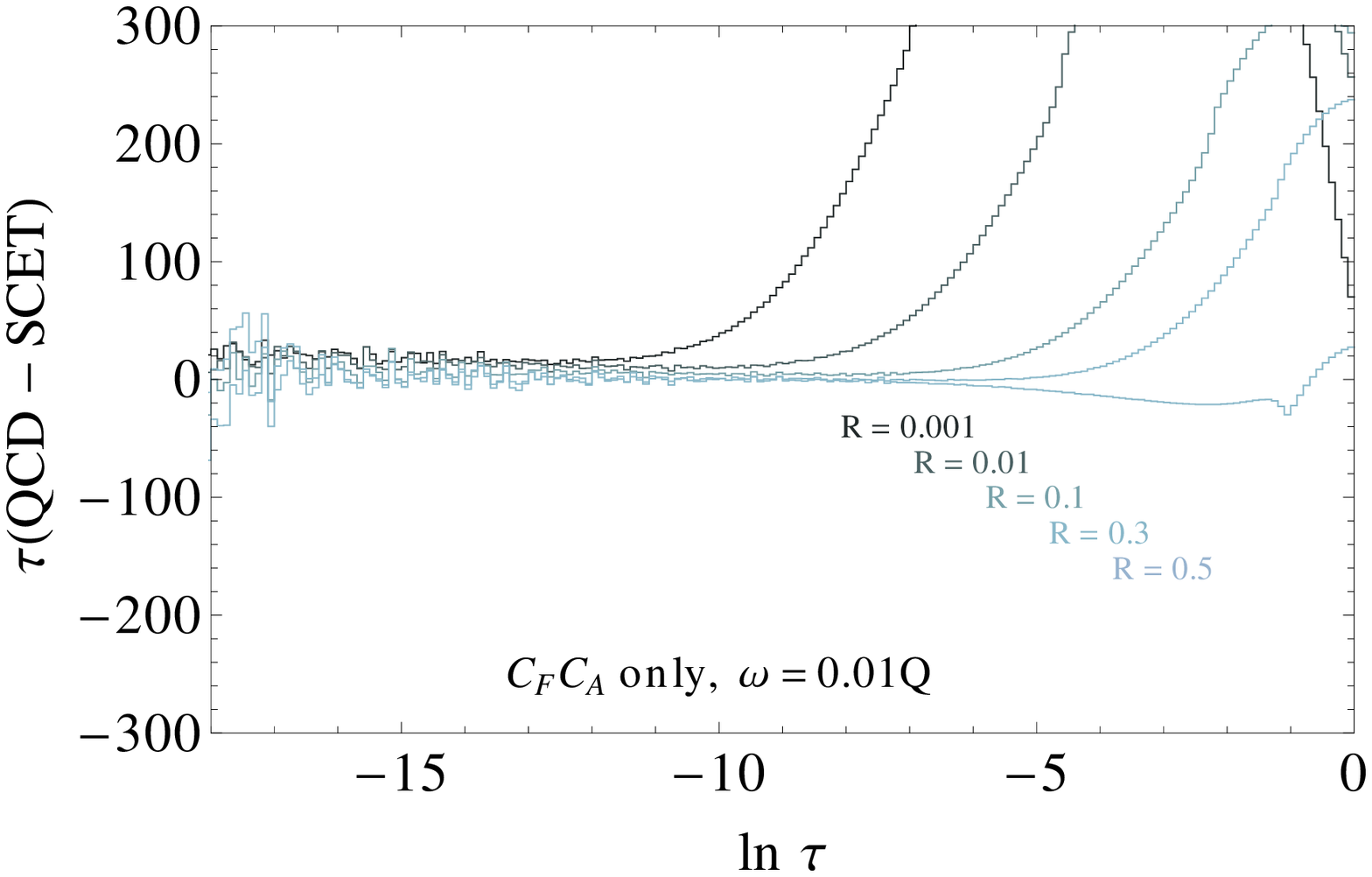}&
  \includegraphics[width=0.5\textwidth]{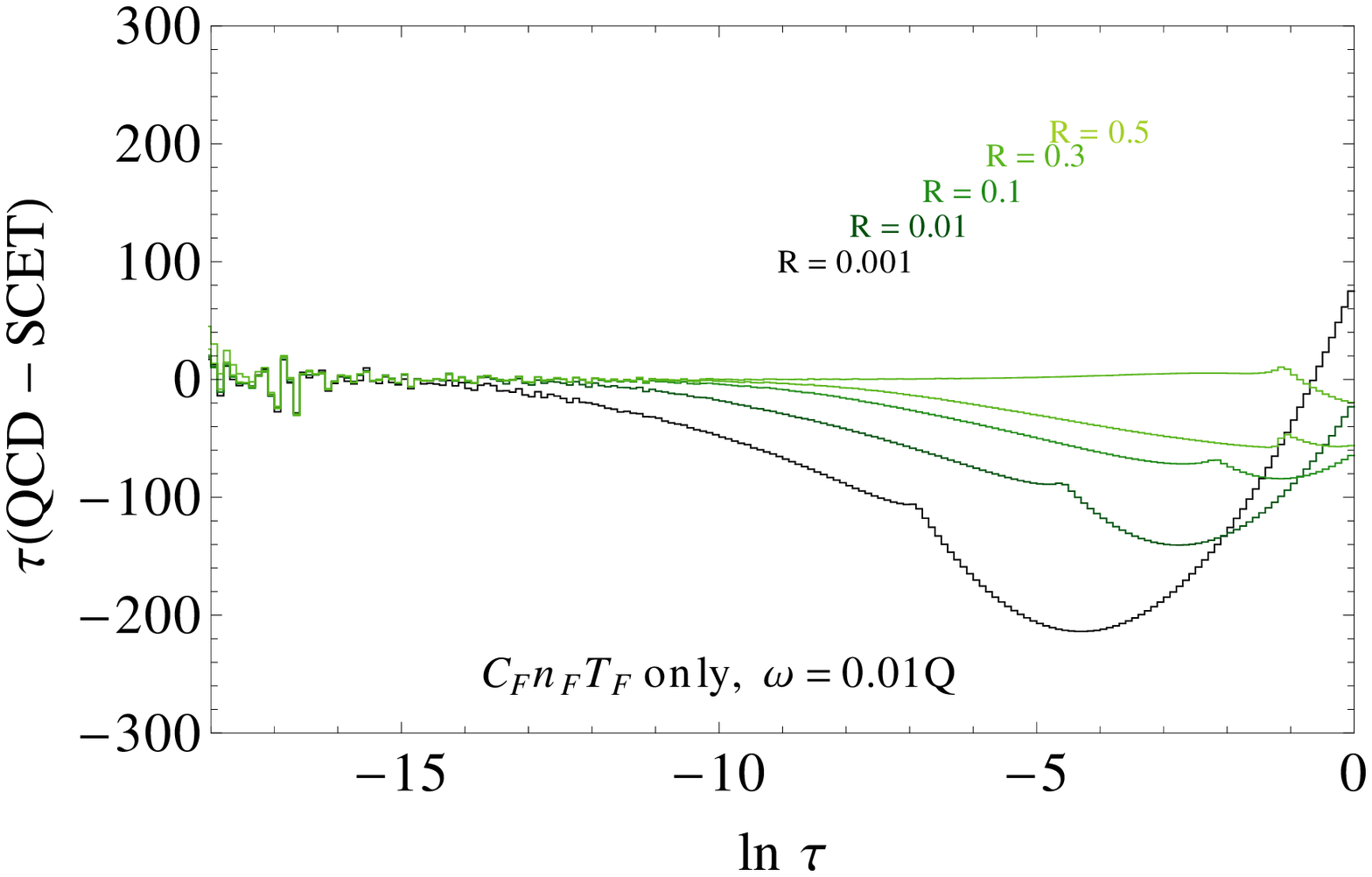}\\
  \includegraphics[width=0.5\textwidth]{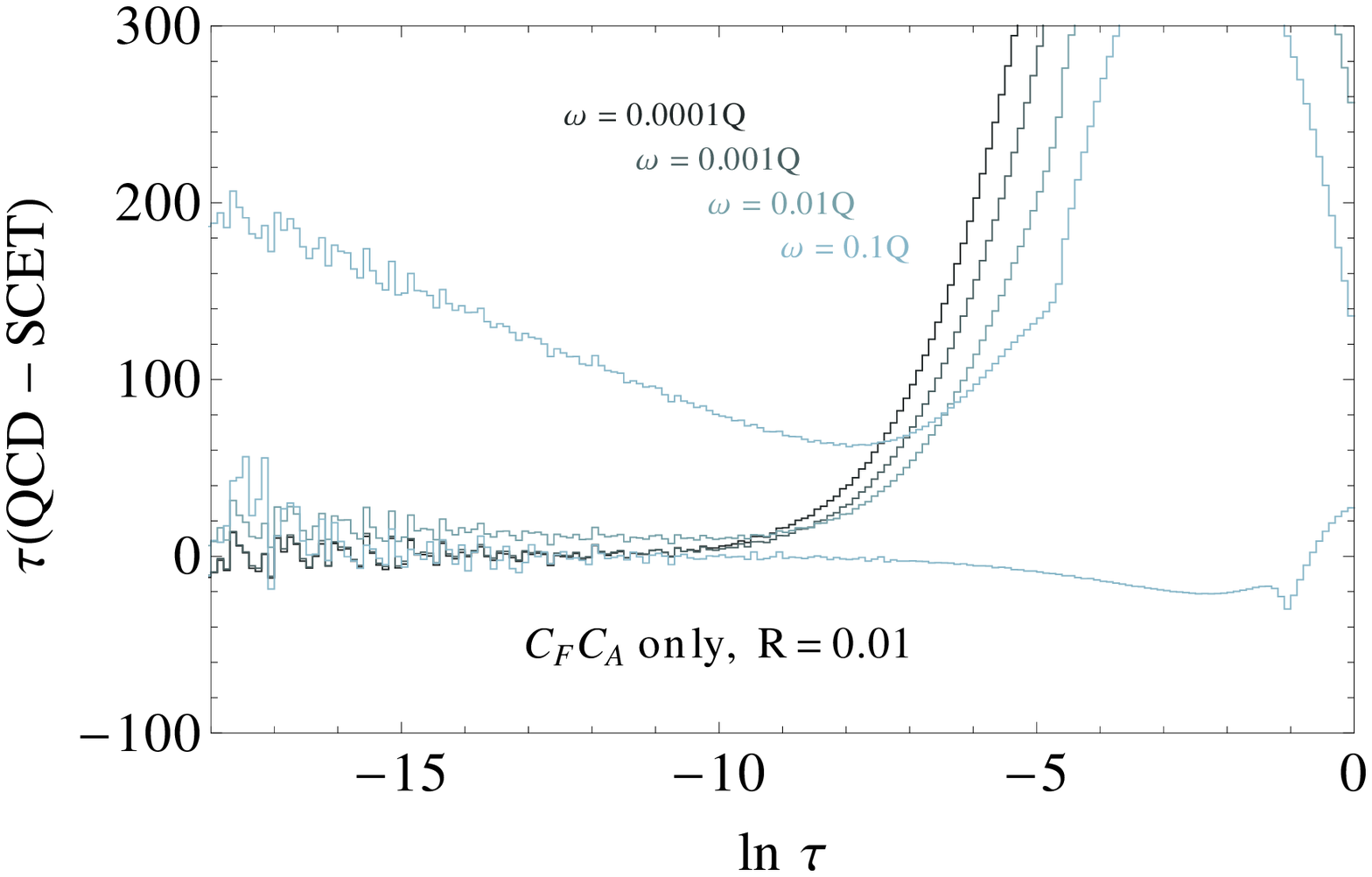}&
  \includegraphics[width=0.5\textwidth]{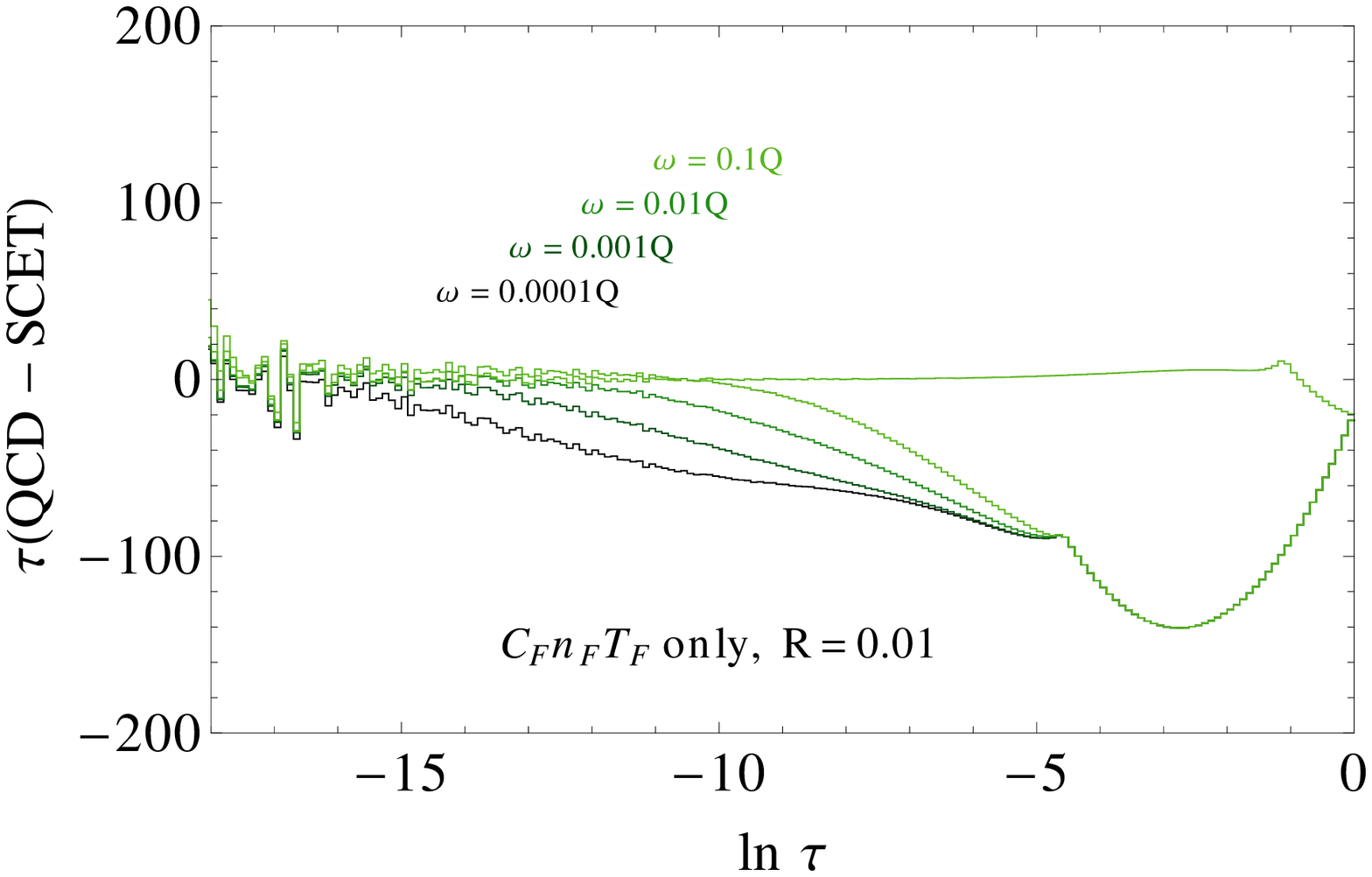}\\
  \end{array}$
\end{center}
  \caption{The difference between the SCET and the QCD calculations of $\frac{\tau_\omega}{\sigma_0}\frac{\dd \sigma}{\dd \tau_\omega}$
for various values of $R$ and $\omega$. The left plots show the $C_F C_A$ color structure and the right plots shows the $C_F n_f T_F$ color structure. The top plots show the variation with $R$ at fixed $\omega = 10^{-2} Q$ and the bottom plots the variation with $\omega$
at fixed $R=10^{-2}$.
For the QCD calculation, a cone algorithm is used with the jet direction taken to be the thrust axis.
}
\label{fig:event2}
\end{figure}

The comparison shown in Figure~\ref{fig:event2} is the difference between the full QCD distribution calculated using {\event} and the two-loop prediction using SCET, with for various values of $R$ and $\omega$. The plots show that SCET is reproducing all of the singular dependence in $\tau_\omega$, for all values of $R$ up to power corrections in $\omega/Q$. This is a non-trivial check that SCET, through factorization formula (\ref{eq:factauo}), is reproducing the correct singular behavior in $\tau_\omega$ for any $R$ and sufficiently small $\omega$.  
Note that one needs $\tom \ll \frac{\omega}{Q}$ to approach the threshold, so for very small $\omega$, the threshold migrates
into a numerically unstable regime, which is apparent in the figure.

\subsection{Analytic checks and definition of the non-global contribution}
\label{sec:diffdef}
Next, we can compare the $\mu$-dependent parts of $K_R^{(2)}(\tau_\omega,\omega,\mu)$ to the $\mu$ dependence predicted by the factorization theorem in SCET (Eq. \eqref{eq:factauo}). With the one- and two-loop anomalous dimensions for the jet and hard functions, the one-loop QCD $\beta$-function coefficient, and $K_R^{(1)}(\tau_\omega,\omega,\mu)$ as inputs, SCET predicts the $\mu$ dependence of the integrated $\tom$ distribution exactly. As it must, the $\mu$ dependence of $K_R^{(2)}(\tau_\omega,\omega,\mu)$ exactly matches the prediction of the factorization theorem.

In order to say more about the non-global contributions to the integrated jet thrust distribution, it is convenient to subtract off the $\mu$-dependent part of $K_R^{(2)}(\tau_\omega,\omega,\mu)$. In fact, as we will argue in Section \ref{sec:smallr}, the preferred way of doing this employs the integrated form of the refactorization ansatz reviewed in Section \ref{sec:review}. In integrated form, the refactorization formula of Eq. \eqref{eq:refact} is
\be
\label{eq:refact2}
K_R(\tom,\omega,\mu) = K^{\in}_R(\tom,\mu)  K^{\out}_R(\omega,\mu) \otimes K^f_R(\tom,\omega)\,.
\ee
In order to make use of the above formula, one also needs the anomalous dimensions associated to $K^{\in}_R(\tom,\mu)$ and $K^{\out}_R(\omega,\mu)$,
\be
\label{eq:regano2}
\gamma^{\rm in}_S = \gamma_S + \gcusp\ln\left(\rr\right) \qquad{\rm and}\qquad\gamma^{\rm out}_S = - \gcusp \ln\left(\rr\right)\,.
\ee
In Eqs. \eqref{eq:regano2} above, $\gamma_S$ is the thrust soft function anomalous dimension. Eq. \eqref{eq:refact2}, Eqs. \eqref{eq:regano2}, and some of the ingredients mentioned in the previous paragraph determine a function $K^{\rm refac}_R (\tau_\omega,\omega,\mu)$ which one can use to define the non-global contributions to the integrated $\tom$ distribution. Explicitly, we have
\bea
&&K^{\rm refac}_R (\tau_\omega,\omega,\mu) \nonumber
 \equiv K^{\in}_R(\tom,\mu)  K^{\out}_R(\omega,\mu) \Big|_{\text{two-loop}} \\
&&= \quad C_F C_A \Bigg[
\ln \left(\frac{\mu }{2 \omega }\right)
 \left(-\frac{176\mathrm{Li}_2(-r)}{3}
-\frac{44}{3} \ln^2(r)-\frac{8}{3} \pi^2 \ln (r)+\frac{536 \ln (r)}{9}-\frac{44 \pi^2}{9}\right)
\el
+\ln \left(\frac{\mu }{\tau_\omega Q}\right)
   \left(-\frac{44}{3} \ln ^2(r)+\frac{8}{3} \pi^2 \ln(r)-\frac{536 \ln (r)}{9}+56 \zeta (3)+\frac{44 \pi^2}{9}-\frac{1616}{27}\right)
\el
+\frac{88}{3} \ln (r) \ln^2\left(\frac{\mu }{2 \omega }\right)
+\left(-\frac{88 \ln (r)}{3}+\frac{8 \pi^2}{3}-\frac{536}{9}\right) \ln ^2\left(\frac{\mu}{\tau_\omega Q}\right)-\frac{176}{9} \ln ^3\left(\frac{\mu}{\tau_\omega Q}\right)
\Bigg]
\el
\qquad + \, C_F n_f T_F \Bigg[
\ln \left(\frac{\mu }{2 \omega }\right) \left(\frac{64
   \mathrm{Li}_2(-r)}{3}+\frac{16 \ln^2(r)}{3}-\frac{160 \ln(r)}{9}+\frac{16 \pi^2}{9}\right)
\el
+ \ln\left(\frac{\mu }{\tau_\omega Q}\right)\left(\frac{16 \ln^2(r)}{3}+\frac{160 \ln(r)}{9}-\frac{16 \pi^2}{9}+\frac{448}{27}\right)
\el
-\frac{32}{3} \ln (r) \ln^2\left(\frac{\mu }{2 \omega}\right)+\left(\frac{32 \ln
   (r)}{3}+\frac{160}{9}\right) \ln ^2\left(\frac{\mu}{\tau_\omega Q}\right)+\frac{64}{9} \ln^3\left(\frac{\mu }{\tau_\omega Q}\right)
\Bigg]\,.
\label{refacansatz}
\eea 

The difference of the exact result and $K^{\rm refac}_R(\tau_\omega,\omega,\mu)$ is
\bea
\label{eq:kdiff}
&&K^{f}_R(\tau_\omega,\omega) \equiv K_R^{(2)}(\tau_\omega,\omega,\mu) - K^{\rm refac}_R(\tau_\omega,\omega,\mu)
\el \qquad\qquad~= 
C_F C_A\Bigg[
\mbox{\large $\chi$}_{C_A}\left({\tau_\omega Q\over 2 \omega},r\right)-\mbox{\large $\chi$}_{C_A}\left(1,r\right)
+\left(16\,\mathrm{Li}_2\left(r^2\right)-\frac{8}{3}\pi^2\right) \ln ^2\left(\frac{\tau_\omega Q}{2 \omega }\right)
\el
+\ln
   \left(\frac{\tau_\omega Q}{2 \omega }\right) \Bigg(-\frac{176
   \mathrm{Li}_2\left(r^2\right)}{3}+8 \mathrm{Li}_3\left(r^2\right)-32
   \mathrm{Li}_3\left(\frac{r^2}{r^2-1}\right)+32 \mathrm{Li}_2\left(r^2\right)
   \ln (r+1)
\el
-64 \mathrm{Li}_2(r) \ln \left(1-r^2\right)-24
   \mathrm{Li}_2\left(r^2\right) \ln \left(r^2\right)+32
   \mathrm{Li}_2\left(r^2\right) \ln \left(1-r^2\right)-64 \mathrm{Li}_3(1-r)
\el
-64
   \mathrm{Li}_3\left(\frac{r}{r+1}\right)-32 \mathrm{Li}_2(-r) \ln (r)-\frac{16
   r^2}{3-3 r^2}+\frac{16}{3} \ln ^3\left(1-r^2\right)-\frac{32 r^2 \ln
   (r)}{3 \left(r^2-1\right)^2}
\el
-\frac{88}{3} \ln \left(r^2\right) \ln
   \left(1-r^2\right)+\frac{32}{3} \ln ^3(r+1)-32 \ln ^2(1-r) \ln
   (r)+\frac{32}{3} \pi ^2 \ln (1-r)
\el
+64 \zeta (3)\Bigg)+\ln
   \left(\frac{\tau_\omega Q}{2 \omega }\right) \left(\frac{16}{3} \pi ^2 \ln (r)-16
   \zeta (3)+\frac{88 \pi ^2}{9}-\frac{8}{3}\right)
\Bigg]
\el
\qquad + C_F n_f T_F \Bigg[
\mbox{\large $\chi$}_{n_f}\left({\tau_\omega Q\over 2 \omega},r\right) - \mbox{\large $\chi$}_{n_f}\left(1,r\right)+\Bigg(\frac{64
   \mathrm{Li}_2\left(r^2\right)}{3}+\frac{32 r^2}{3
   \left(1-r^2\right)}+\frac{64 r^2 \ln (r)}{3\left(1-r^2\right)^2}
\el
+\frac{64}{3} \ln (r) \ln \left(1-r^2\right)\Bigg)
   \ln \left(\frac{\tau_\omega Q}{2 \omega }\right)+\left(\frac{16}{3}-\frac{32 \pi^2}{9}\right)
 \ln \left(\frac{\tau_\omega Q}{2 \omega }\right)
\Bigg]\,.
\eea
As required, this difference 
is $\mu$-independent and contains all of the non-global logarithms. The functions $\mbox{\large $\chi$}_{C_A}(x,r)$ and $\mbox{\large $\chi$}_{n_f}(x,r)$ in Eq. \eqref{eq:kdiff} are defined in Appendix~\ref{app:chi}. There should
be an additional $\tom$- and $\omega$-independent function of $R$ in Eq. \eqref{eq:kdiff}, which has not
been calculated. As mentioned above, this function would not affect the differential $\tom$ distribution
and is therefore of secondary importance.

In recent years, quite a bit of effort has been devoted to understanding the structure of NGLs. So far, most of the analytical calculations have focused on the leading logarithms. For example, in Eq. (\ref{eq:kdiff}) above, the leading NGL is simply 
\be
C_F C_A\left(16\,\mathrm{Li}_2(r^2)-\frac{8}{3}\pi^2\right)\ln^2 \left(\frac{\tau_\omega Q}{2 \omega }\right)\,.
\ee
which agrees with results from Refs.~\cite{Kelley:2011tj,Hornig:2011tg}.
Leading NGLs are certainly of interest but such computations are not, by themselves, likely to be of much use if our ultimate goal is to understand the resummation of NGLs. On the other hand, our exact result for the non-global contribution to the integrated $\tau_\omega$ distribution is extremely complicated (both $\mbox{\large $\chi$}_{C_A}\left(x,r\right)$ and $\mbox{\large $\chi$}_{n_f}\left(x,r\right)$ depend on two independent scales in a highly non-trivial way) and this makes it difficult to gain any further theoretical insight. If there is any hidden structure in our results, it should be significantly easier to identify in the small $R$ limit.

\section{The Small $R$ Limit of the Non-Global Logarithms}
\label{sec:smallr}
It was observed in Ref.~\cite{Kelley:2011tj} that, in the small $R$ limit, the jet thrust soft function simplifies. 
Having computed the exact scale-dependent contributions to the integrated jet thrust distribution at finite $R$, we can now understand more precisely what is happening. As we will see, in the small $R$ limit, the non-global contribution to the integrated $\tau_\omega$ distribution is determined completely by the non-global contribution to the integrated hemisphere soft function.

It has been noted by a number of authors that, for jet observables defined with the anti-$k_T$ jet algorithm, one typically finds that the $R \rightarrow 0$ limit of the leading NGL is simply given by the result in the hemisphere case, up to possibly a factor of two depending on the observable in question~\cite{Banfi:2010pa,Kelley:2011tj,Hornig:2011iu,Hornig:2011tg}. The jet algorithm employed in this paper is very closely related to the anti-$k_T$ algorithm. In fact, the in-in and in-out contributions are exactly the same in both jet algorithms. The only difference between the algorithms at two loops is that they treat the out-of-jet radiation differently. We strongly suspect that the NGLs in the two jet algorithms actually coincide, at least in the small $R$ limit. This implies that we should be able to check whether, for the integrated anti-$k_T$ jet thrust distribution, one can relate the next-to-leading NGLs to analogous NGLs in the integrated hemisphere soft function calculated in Refs.~\cite{Kelley:2011tj} and \cite{Hornig:2011iu}; since we have computed the exact expression, we are now in a position to go beyond the leading-logarithm approximation. In what follows, we assume some familiarity with the conventions and definitions that we established in Section \ref{sec:calculation}.

\subsection{In-in rescaling}
\label{sec:ininanalysis}
Before discussing the in-out contributions and non-global logarithms, it is instructive to first consider the simpler case where both soft particles get clustered into the same jet. The same-side in-in contribution to the integrated $\tau_\omega$ distribution is
\begin{eqnarray}
\label{eq:iiintegrand1}
K^{r_1}_R(\tau_\omega,\omega,\mu)
&=&
2 \int_0^{\tau_\omega}\dd \tau_\omega^\prime \int_0^\omega \dd \lambda \int_0^\infty \dd k_L \dd k_R \,(4 \pi)^4 \left( \frac{\mu^2 e^{\gamma_E}}{4\pi}\right)^{2\ep}
\int\! \frac{\dd^d q}{(2\pi)^d}
\frac{\dd^d k}{(2\pi)^d}
\el
\times \left(\frac{1}{2!}I_{C_A} + I_{n_f}\right) (-2\pi i)^2 \delta(q^2) \Theta\left(q^0\right)\delta(k^2)\Theta\left(k^0\right) \Theta(r k^- - k^+) 
\el
\times \Theta(r q^- - q^+) \delta(k_L) \delta(k_R - q^+ - k^+) \delta\left(\lambda\right) \delta\left(\tau_\omega^\prime - {k_L + k_R\over Q}\right)\,,
\end{eqnarray}
where $I_{C_A}$ and $I_{n_f}$ are the two-loop, two-parton integrands for the $C_F C_A$ and $C_F n_f T_F$ color structures, given respectively by Eqs. (16) and (27) of Ref. \cite{Kelley:2011ng}.

\begin{figure}[t]
\begin{center}
  \includegraphics[width=0.7\textwidth]{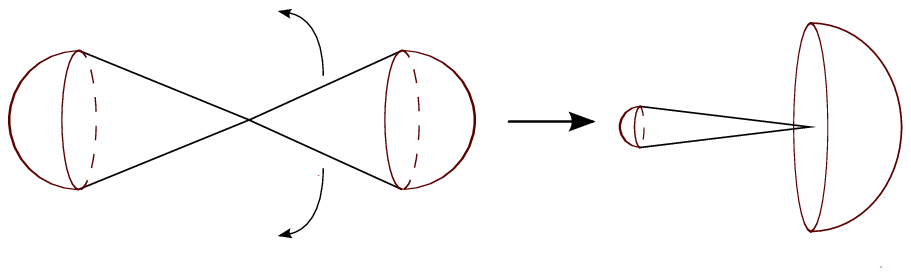}
  \caption{By rescaling one of the jet directions a small cone can be mapped to a hemisphere. Rescaling in this way makes some universal properties of non-global logarithms easier to identify.}
\end{center}
\label{fig:cone}
\end{figure}

To simplify the theta and delta functions, we can rescale the ${\bf n}$-jet Wilson line. This turns the cone jet into a hemisphere jet, as illustrated in Figure~\ref{fig:cone}. At the level of the integrand, this rescaling amounts to making the replacements:
\begin{equation}
  k^- \to \frac{~k^-}{r} \qquad {\rm and} \qquad q^- \to {~q^- \over r}\,.
\end{equation}
This change of variables induces the transformations
\begin{equation}
I_{C_A} \to r^2 I_{C_A} \qquad {\rm and} \qquad
I_{n_f} \to r^2 I_{n_f}
\end{equation}
and the phase-space cuts in Eq. (\ref{eq:iiintegrand1}) that depend on the minus components of the light-cone momenta become
\begin{equation} 
 \Theta(k^- - k^+) \Theta(q^- -  q^+).
\end{equation}
Remarkably, the transformed integrand reduces to the $\Ord\left(\alpha_s^2\right)$ same-side hemisphere integrand times an overall factor of $r^{2\e}$.
A similar rescaling was observed in the context of beam functions in~\cite{Stewart:2009yx}.

 After performing the trivial integrations over $k_L$ and $k_R$, the result is simply expressed in terms of the same-side contributions to the integrated hemisphere soft function:
\bea
\label{eq:iiintegrand2}
K^{r_1}_{R}(\tau_\omega,\omega,\mu)
&=& 
2 r^{2 \ep}
\int_0^{\tau_\omega} Q \,\dd \tau_\omega^\prime \int_0^\omega \dd \lambda
\frac{\delta(\lambda) \mu^{4\ep}}{\left(Q \,\tau_\omega^\prime\right)^{1+4\ep}}
\el
~~~~~~\times\Bigg[ C_F C_A \mbox{\Large $g$}^{\rm hemi}_{C_A}\left(\ep\right) + C_F n_f T_F \mbox{\Large $g$}^{\rm hemi}_{n_f}\left(\ep\right)\Bigg]\,,
\eea
where $\mbox{\Large $g$}^{\rm hemi}_{C_A}(\ep)$ and $\mbox{\Large $g$}^{\rm hemi}_{n_f}(\ep)$ are the same-side contributions to the NLO hemisphere soft function, given in Eqs. (18) and (29) of Ref.~\cite{Kelley:2011ng}.

We used this rescaling to compute the two-loop same-side in-in integrals in Section~\ref{sec:calculation}. In fact, this result is straightforward to generalize. At any order in perturbation theory, if all of the particles go into a single jet, the contribution to the integrated $\tau_\omega$ soft function will be given by the analogous hemisphere calculation with the light-cone momenta appropriately rescaled in the manner described above.

Before moving on, we briefly consider the case where the two partons end up in separate jets and show that this contribution is suppressed in the small $R$ limit. First, for any $r$, this contribution to $K_R^{(2)}(\tau_\omega,\omega,\mu)$ is given by
\begin{eqnarray}
\label{eq:iiintegrand3}
K^{r_2}_R(\tau_\omega,\omega,\mu)
&=&
2 \int_0^{\tau_\omega}\dd \tau_\omega^\prime \int_0^\omega \dd \lambda \int_0^\infty 
\dd k_L \dd k_R \,
(4 \pi)^4 \left( \frac{\mu^2 e^{\gamma_E}}{4\pi}\right)^{2\ep}
\int\! \frac{\dd^d q}{(2\pi)^d}
\frac{\dd^d k}{(2\pi)^d}
\el
\times \left(\frac{1}{2!}I_{C_A} + I_{n_f}\right) (-2\pi i)^2 \delta(q^2) 
\Theta\left(q^0\right)\delta(k^2)\Theta\left(k^0\right) 
\Theta(r k^- - k^+)  
\el
\times \Theta(r q^+ - q^-)\delta(k_R - k^+)\delta(k_L - q^-) 
\delta\left(\lambda\right) \delta\left(\tau_\omega^\prime - {k_L + k_R\over Q}\right)\,.
\end{eqnarray}
Let us try to proceed as above, this time rescaling $k^-$ and $q^+$:
\begin{equation}
\label{eq:oppininresc}
  k^- \to \frac{~k^-}{r} \qquad {\rm and} \qquad q^+ \to {~q^+ \over r}\,.
\end{equation}
This time $I_{C_A}$ and $I_{n_f}$ do not transform simply under the rescaling. Instead we find that 
\begin{equation}
I_{C_A} \to r^3 \tilde{I}_{C_A}(r) \qquad {\rm and} \qquad
I_{n_f} \to r^5 \tilde{I}_{n_f}(r)\,,
\end{equation}
where $\tilde{I}_{C_A}(r)$ and $\tilde{I}_{n_f}(r)$ are regular but non-vanishing at $r = 0$. Since, as before, the integration measure contributes a factor $r^{-2 + 2\ep},$ we see that this entire contribution is suppressed at small $R$ relative to the one studied above. In fact, in the small $R$ limit, the opposite-side in-in contributions are suppressed relative to all of the other contributions and can therefore be neglected.

\subsection{In-out rescaling at small $R$}
\label{sec:inoutanalysis}
Next, we will look at the contributions where one soft parton gets clustered into a jet and the other does not. Here we will find a similar mapping to the hemisphere integrals but only at small $R$. At the level of the integrand, the in-out contribution to the integrated $\tau_\omega$ distribution is
\cmb{-1 cm}{0 cm}
\bea
\label{eq:iointegrand1}
K^{r_3}_R(\tau_\omega,\omega,\mu)
&=&
4 \int_0^{\tau_\omega}\dd \tau_\omega^\prime \int_0^\omega \dd \lambda \int_0^\infty \dd k_L \dd k_R \,(4\pi)^4 \left( \frac{\mu^2 e^{\gamma_E}}{4\pi}\right)^{2\ep}
\int\! \frac{\dd^d q}{(2\pi)^d}
\frac{\dd^d k}{(2\pi)^d} (-2\pi i)^2
\el
\times \left(\frac{1}{2!}I_{C_A} + I_{n_f}\right) \delta(q^2) \Theta\left(q^0\right)\delta(k^2)\Theta\left(k^0\right) \Theta(r k^- - k^+) \Theta(q^- - r q^+) 
\el
\times \Theta(q^+ - r q^-) \delta(k_R - k^+)\delta\left(\lambda - \frac{q^+ + q^-}{2}\right) \delta(k_L) \delta\left(\tau_\omega^\prime - {k_L + k_R\over Q}\right)\,.
\eea
\cme

Changing variables in an attempt to map the cone jet to a hemisphere jet as above, we see that the phase-space cuts in Eq. (\ref{eq:iointegrand1}) that depend on the minus components of the light-cone momenta can be written as 
\begin{equation}
2 r \Theta(k^- - k^+) \Theta(q^- - r^2 q^+)
\Theta(q^+ - q^-)
\delta\left(q^+ r + q^- - 2 r \lambda\right)\,.
\end{equation}
This is not exactly a hemisphere projection for general $R$. However, in the small $R$ regime, the above expression simplifies to
\begin{equation}
2 r \,\Theta(k^- - k^+)\Theta(q^-)
\Theta(q^+ - q^-)
\delta\left(q^- - 2 r \lambda\right)\,. 
\label{eq:hemiform}
\end{equation}

The phase-space cuts enforced by (\ref{eq:hemiform}) are identical to the ones which arise in the calculation of the opposite-side contributions to the two-loop hemisphere soft function, provided that one makes the replacement $k_L \rightarrow 2 r \lambda$. To see this, after performing the trivial integrations over $k_L$ and $k_R$ in Eq. (\ref{eq:iointegrand1}) above, replace the phase-spaces cuts that depend on $k^-$ and $q^-$ using (\ref{eq:hemiform}) and then compare the resulting product of thetas and deltas in the integrand to the second line of Eq. (13) in Ref. \cite{Kelley:2011ng} (with $k_L = 2 r \lambda$ and $k_R = Q \tom^\prime$). With this understanding, we find that the above expression can be written in terms of the opposite-side contributions to the integrated hemisphere soft function:
\bea
\label{eq:iointegrand}
K^{r_3}_{R \to 0}(\tau_\omega,\omega,\mu)
&=& 
2 r^{2 \ep} \left(\frac{\als}{4\pi}\right)^2
\int_0^{\tau_\omega Q} \dd x \int_0^{2\, r \,\omega} \dd y
\frac{\mu^{4\ep}}{(x y)^{1+2\ep}}
\el
~~~~\times
\Bigg[ C_F C_A \mbox{\Large $f$}^{\rm hemi}_{C_A}\left(\frac{x}{y},\ep\right)+C_F n_f T_F \mbox{\Large $f$}^{\rm hemi}_{n_f}\left(\frac{x}{y},\ep\right)\Bigg]\,,
\eea
where $\mbox{\Large $f$}^{\rm hemi}_{C_A}(z, \ep)$ and $\mbox{\Large $f$}^{\rm hemi}_{n_f}(z, \ep)$ are the opposite-side contributions to the NLO hemisphere soft function, defined in Eqs. (17) and (28) and Appendix A of Ref.~\cite{Kelley:2011ng}. In deriving Eq. (\ref{eq:iointegrand}) we found it useful to make the change of variables
\begin{equation}
\tau_\omega^\prime = {x \over Q} \qquad\qquad \lambda = {y \over 2 r}\,.
\end{equation}

This correspondence between the in-out contributions to the integrated $\tau_\omega$ distribution and the opposite-side contributions to the integrated hemisphere soft function is valid for sufficiently small $R$ and arbitrary values of $\tau_\omega$ and $\omega$. The striking similarity between Eqs. (\ref{eq:iiintegrand2}) and (\ref{eq:iointegrand}) is suggestive. In fact, for sufficiently small $R$, we will show in the next section that it is possible to very accurately model {\it all} of the terms in the integrated $\tau_\omega$ distribution not captured by the refactorization ansatz -- including all of the NGLs -- using the non-global contributions to the integrated hemisphere soft function.

\subsection{Small $R$ limit}
\label{sec:smallRresult}
We have shown that the in-out contributions at small $R$ (and so $r\sim R$) are given by the opposite-side contributions to the integrated hemisphere soft function times an overall factor of $2 R^{2\e}$. We were inspired by this small $R$ correspondence to take a closer look at the non-global contributions to the integrated hemisphere soft function, $\mathcal{R}_f\left(z\right)$. 

Using the results of Sections \ref{sec:ininanalysis} and \ref{sec:inoutanalysis}, we 
are able to take the small $R$ limit of 
the scale-dependent non-global contribution to the integrated $\tom$ distribution that we presented in Eq.\eqref{eq:kdiff}.
We find
\begin{equation}
\label{eq:simpkdiff}
K^{f}_{R\to 0}(\tau_\omega,\omega) 
= 2\Bigg[\mathcal{R}_f\left({\tau_\omega Q \over 2 R \omega}\right)-\mathcal{R}_f\left({1\over R}\right)\Bigg],
\end{equation}
where $\mathcal{R}_f(z)$ is the non-global contribution to the integrated hemisphere soft function, defined in Eq. (53) of Ref. \cite{Kelley:2011ng}:
\begin{eqnarray}
\label{sf:exact}
&&\mathcal{R}_f(z) =  
\left[
   -88 \text{Li}_3(-z)-16 \text{Li}_4\left(\frac{1}{z+1}\right)-16
   \text{Li}_4\left(\frac{z}{z+1}\right)+16 \text{Li}_3(-z) \ln
   (z+1)
\right.
\nn
&& 
\left.
   +\frac{88 \text{Li}_2(-z) \ln (z)}{3}-8 \text{Li}_3(-z) \ln
   (z)-16 \zeta (3) \ln (z+1)+8 \zeta (3) \ln (z)-\frac{4}{3} \ln^4(z+1)
\right.
\nn
&&
\left.
   +\frac{8}{3} \ln (z) \ln^3(z+1)+\frac{4}{3} \pi^2 \ln^2(z+1)-\frac{4}{3} \pi^2 \ln^2(z)-\frac{4 \left(3 (z-1)+11 \pi^2 (z+1)\right) \ln (z)}{9 (z+1)}
\right.
\nn
&& 
\left.
    -\frac{506 \zeta (3)}{9}+\frac{16\pi^4}{9}-\frac{871 \pi^2}{54}-\frac{2032}{81}\right]C_F C_A
+
  \left[ 
  32 \text{Li}_3(-z)-\frac{32}{3} \text{Li}_2(-z) \ln (z)
\right.
\nn
&& 
\left.
    +\frac{8 (z-1) \ln (z)}{3 (z+1)}+\frac{16}{9} \pi^2 \ln (z)+\frac{184 \zeta (3)}{9}+\frac{154 \pi^2}{27}-\frac{136}{81}
  \right]
  C_F n_f T_F\,.
\end{eqnarray}
In deriving this small $R$ limit, the rescaling of the in-out integrals in particular was critical -- it is certainly not straightforward to derive Eq. (\ref{eq:simpkdiff}) directly from the expression for $K^f_R(\tau_\omega,\omega)$ given in Eq. (\ref{eq:kdiff}).
Again, there may be additional contributions to $K^f_{R\to0}(\tom,\omega)$ which are $\tom$- and $\omega$-independent and have not been computed.

\begin{figure}[t]
  \includegraphics[width=0.8\textwidth]{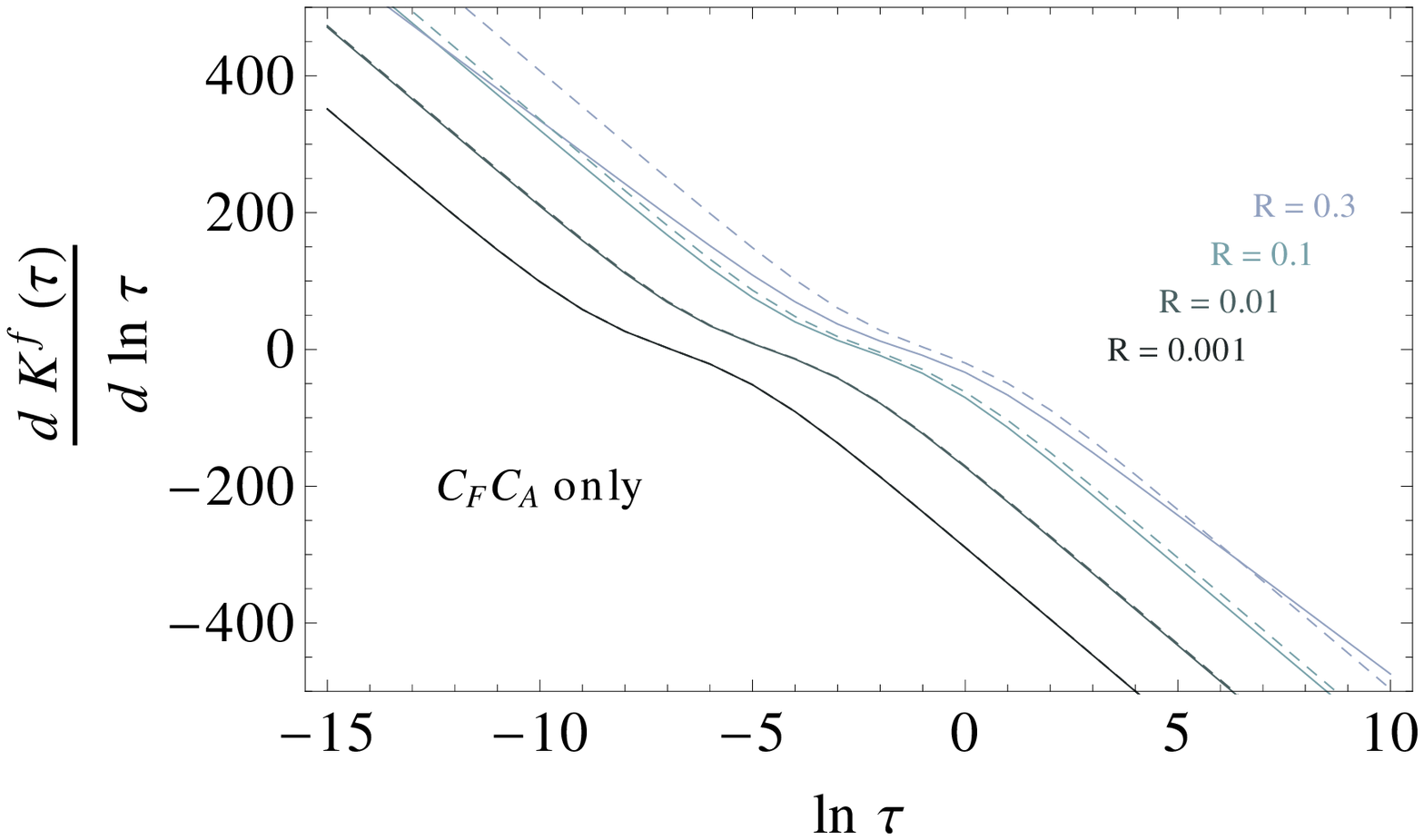}
  \includegraphics[width=0.8\textwidth]{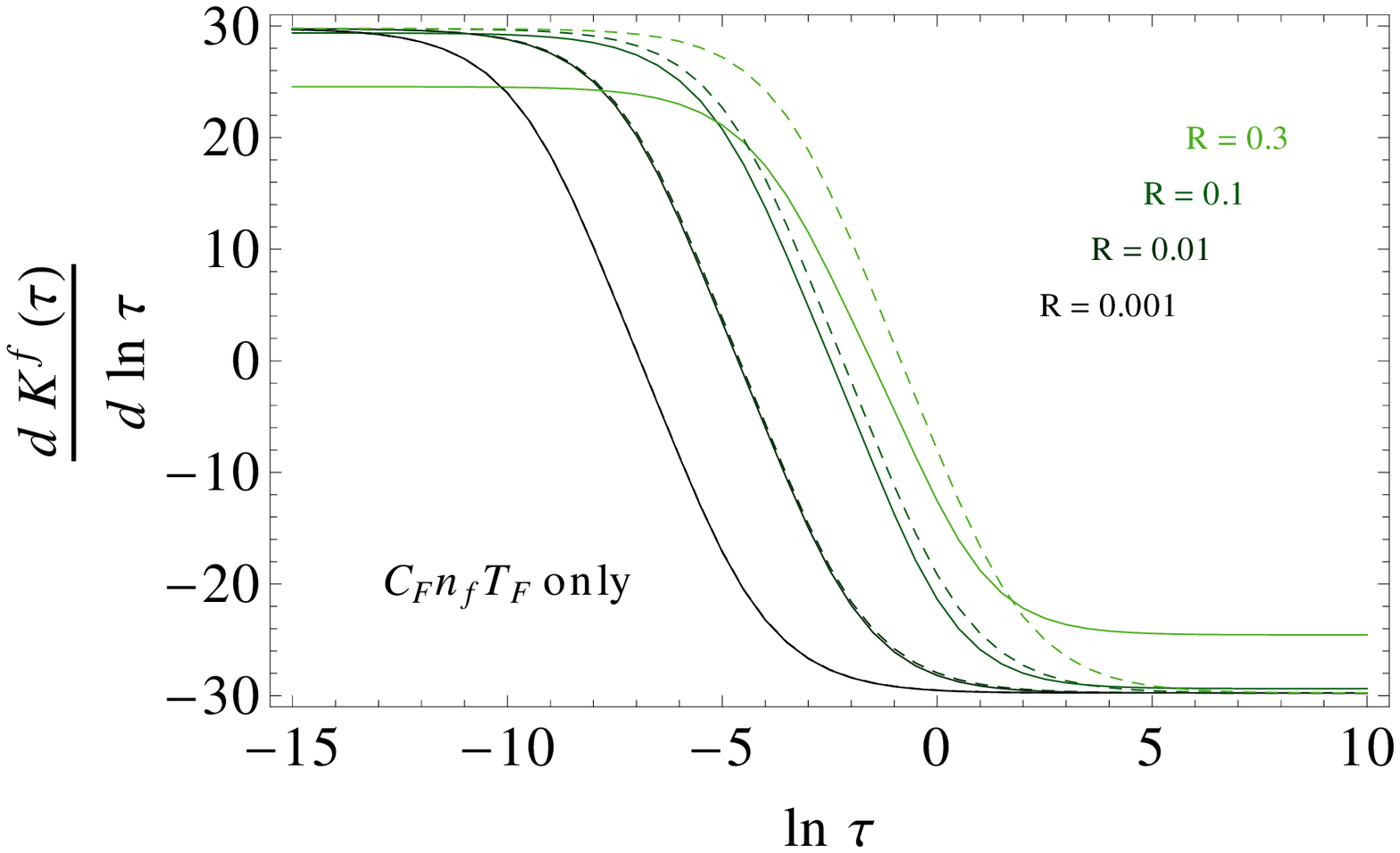}
  \caption{Contributions to the differential jet thrust distribution coming from the non-global terms in the soft function. 
We show ${\dd\over \dd \ln (\tau)} K_R^f(\tau)$ (solid) and its small $R$ limit,
 $\frac{\dd}{\dd \ln \tau}K^{f}_{R\to 0}(\tau)$ (dashed),
plotted as a function of $\tau = {\tom Q \over 2 \omega}$ for various $R$.}
\label{fig:ngl}
\end{figure}

In Figure~\ref{fig:ngl}, we compare the $\tom$ distributions coming from the exact expression, $K^f_R(\tau_\omega,\omega)$, and from
the small $R$ approximation, 
$K^f_{R\to 0}(\tau_\omega,\omega)$, for both non-trivial color structures. We see that the approximation
works spectacularly well and remains valid for small but experimentally relevant values of the jet cone size $\left(R \sim 0.1\right)$. 

\subsection{Extraction of the non-global logarithms}
\label{sec:nglextract}
Traditionally, one would define the non-global logarithms as those terms in $K^f_R(\tau_\omega, \omega)$ which diverge in the $\left| \ln\left(\frac{\tau_\omega Q}{2\omega}\right)\right|\gg 1$ limit. However, the non-global logarithms in our integrated jet mass distribution are somewhat more subtle since there is a term in $K^f_R(\tau_\omega, \omega)$ which diverges in the $R \to 0$ limit as well. To sensibly extract the NGLs in our case, we actually need to consider the double limits $1 \gg R \gg \frac{\tau_\omega Q}{2\omega}$ and $\frac{\tau_\omega Q}{2\omega} \gg 1 \gg R$. We found in Section~\ref{sec:smallRresult} that
\begin{equation}
\label{eq:OR}
K^{f}_{R}(\tau_\omega,\omega) 
= 2 \mathcal{R}_f\left({\tau_\omega Q \over 2 R \omega}\right)+ f(R) + \mathcal{O}(R) \,,
\end{equation}
where $f(R)$ is a $\omega$-independent and $\tom$-independent function of $R$ which we have not calculated.
The appearance of the function $\mathcal{R}_f(z)$ in Eq.~(\ref{eq:OR}) suggests that the leading and next-to-leading NGLs in the integrated $\tom$ distribution ought to be simply related to the leading and next-to-leading hemisphere NGLs derived in Refs.~\cite{Kelley:2011ng,Hornig:2011iu}.

We can easily extract the NGLs in the integrated $\tom$ distribution by first taking the $\left|\ln\left(\frac{\tau_\omega Q}{2 \omega}\right)\right|\gg 1$ limit of the non-global function $K^f_{R}(\tom, \omega)$ and then taking the small $R$ limit of the expression that results. In this approximation, we find that the function $K^f_{R}(\tom, \omega)$ reduces to
\begin{eqnarray}
\label{eq:asympt}
K^{f,\,\text{NGL}}_{R \to 0}(\tom,\omega) &=& 
C_F C_A\Bigg[
-\frac{8\pi^2}{3}
\ln^2\left(\frac{\tau_\omega Q}{2 R \omega}\right)
+ \Bigg(
-\frac{8}{3} + \frac{88\pi^2}{9} -16\zeta_3\Bigg)
\left|\ln\left(\frac{\tau_\omega Q}{2 R \omega}\right)\right|
\Bigg]
\el
+ C_F n_f T_F
\Bigg(
\frac{16}{3} - \frac{32\pi^2}{9}\Bigg)
\left|\ln\left(\frac{\tau_\omega Q}{2 R \omega}\right)\right| + \cdots \,.
\end{eqnarray}
Strictly speaking, Eq. (\ref{eq:asympt}) is only correct up to terms independent of $\tom$ and $\omega$ ({\it i.e.} functions unconstrained by the calculations performed in this paper). However, although it remains to be conclusively proven, we strongly suspect that Eq. (\ref{eq:asympt}) actually captures {\it all} of the singular $R$ dependence present in the contributions to the integrated $\tau_\omega$ distribution not determined by the refactorization ansatz. After multiplying the above expression by $1/2$ and making the substitution $\ln\left(\frac{\tau_\omega Q}{2R\omega}\right) \rightarrow \ln(z)$, we find that Eq. (\ref{eq:asympt}) exactly reproduces the hemisphere NGLs of Eq.~(54) in Ref.~\cite{Kelley:2011ng}. This shows that, as hoped, the next-to-leading NGLs in our integrated jet thrust distribution are simply related to corresponding next-to-leading NGLs in the integrated hemisphere soft function. In fact, if it turns out that, as seems likely, the anti-$k_T$ NGLs coincide with ours in the small $R$ limit, we have shown that there is no essential difference between the small $R$ NGLs that arise in the integrated anti-$k_T$ jet thrust distribution and the NGLs that arise in the hemisphere mass distribution.
%

Clearly, the simplicity of the NGLs derived above suggests that a simple factorized picture emerges in the small $R$ limit. It seems likely that, for sufficiently small $R$, the $\tom$ soft function refactorizes to all orders in perturbation theory: 
\begin{eqnarray}
\label{eq:refinedrefac}
S_R(k_L,k_R,\lambda, \mu)
&=&
S_\mu\left(\frac{k_L}{\sqrt{R} \mu}\right)
S_\mu\left(\frac{k_R}{\sqrt{R} \mu}\right)
S_{\rm out}\left(\frac{2 \sqrt{R}\lambda}{\mu},R\right)
\el
~~~~~~~~~~~~~~~~~~~~~\otimes\,
S_f\left(\frac{k_L}{2 R\lambda}\right)
S_f\left(\frac{k_R}{2 R\lambda}\right)
+ \mathcal{O}(R).
\end{eqnarray}
Here, the functions $S_\mu(z)$ and $S_f(z)$ denote respectively the $\mu$-dependent and $\mu$-independent parts of the hemisphere soft function~\cite{Chien:2010kc,Kelley:2011ng,Hoang:2008fs}. Writing the soft function in this refactorized form makes the relationship between the small $R$ exclusive jet mass soft function and the hemisphere soft function completely explicit.
It is worth pointing out that $S_\mu(z)$ and $S_f(z)$ have no explicit $R$ dependence; their $R$ dependence is simply encoded in their arguments, 
$\frac{k}{\sqrt{R}\mu}$ and $\frac{k}{2 R \lambda}$. In fact, the only part of \eqref{eq:refinedrefac} which is not a simple 
rescaling of part of the hemisphere soft function
is $S_{\rm out}\left(\frac{2 \sqrt{R}\lambda}{\mu},R\right)$. Moreover, the dependence on the first argument of 
$S_{\rm out}\left(\frac{2 \sqrt{R}\lambda}{\mu},R\right)$ is fixed by RG invariance.
Actually, it is possible to fix the entire singular $R$ dependence of $S_{\rm out}(z,R)$  with arguments similar to the ones made in Sections~\ref{sec:ininanalysis} and \ref{sec:inoutanalysis}. This analysis, however, is beyond the scope of the current paper.

When we expand Eq.~\eqref{eq:refinedrefac} to $\Ord\left(\alpha_s^2\right)$, we find that
\begin{equation}
K^{f}_{R}(\tau_\omega,\omega) 
= 2 \mathcal{R}_f\left({\tau_\omega Q \over 2 R \omega}\right) + \mathcal{O}(R) \,.
\end{equation}
This is completely consistent with the function $K_{R \to 0}^f(\tom,\omega)$, Eq.~\eqref{eq:simpkdiff}, coming from the expansion of the exact result. It also predicts that the unknown $\tom$- and $\omega$-independent function of $R$ in the integrated distribution should vanish
in the small $R$ limit.
Furthermore, since we have confirmed this refactorization at two loops, including non-global logs, the refactorization and $\Gamma_{\mathrm{cusp}}$ ansatz
of Ref.~\cite{Kelley:2011tj} are also confirmed at two loops -- Eq.~\eqref{eq:refinedrefac} reproduces them when $S_f(z)=1$. 


\section{Conclusions}
\label{sec:conclusions}
Over the last several years it has become increasingly apparent that jet substructure will play an important role
in collider physics, and that theoretical calculations of jet substructure will require new organizational
tools. For example, the simplest substructure observable, jet mass, when examined in an exclusive context, such as for the hardest or two
hardest jets, is associated with poorly understood dependence on the multiple relevant scales (jet masses and veto scales).
 In this paper, by studying in detail the non-global structure of the soft, scale-dependent contributions to the integrated cone jet thrust distribution at two-loops, we took an important step towards understanding the interplay of these scales. 
We presented compelling evidence that the non-global logarithms that arise in a realistic class of jet algorithms when
expanded around small jet size $R$
are in fact no more complicated than the non-global logarithms that arise when one uses the hemisphere jet algorithm.

It seems that at each order in perturbation theory, it is possible to take the limit of small jet cone size in a consistent and useful way at the level of the integrands and that taking this limit effectively reduces all subsequent integrations to the corresponding hemisphere ones. 
This vastly simplifies the non-global parts of the distribution.
A comparison of the non-global contribution to the integrated hemisphere mass distribution given in Section \ref{sec:smallRresult} and the analogous integrated jet thrust distribution given in \ref{sec:diffdef} illustrates the point -- the hemisphere results of Ref.~\cite{Kelley:2011ng} reproduced in Eq. \eqref{sf:exact} are far simpler than the finite $R$ results in Eq. \eqref{eq:kdiff}, derived in this paper. Na\"{i}vely, one might think that this procedure, explained in more detail in Sections \ref{sec:ininanalysis} and \ref{sec:inoutanalysis}, only gives an accurate approximation if one works with unrealistically small jet resolution parameters. However, as illustrated in Figure \ref{fig:ngl}, this is not the case at all; we found that approximating our result for the finite $R$ differential distribution using this procedure works great even for  $R\sim 0.1$.

One point worth emphasizing is that the refactorization ansatz of Ref.~\cite{Kelley:2011tj} played an important role in our analysis. In Figure \ref{fig:refact} we compared the $\Ord\left(\als^2\right)$ prediction of this ansatz to full QCD at the same order using {\event} and showed that the refactorization ansatz reproduces the singular terms in the QCD differential distribution very nicely, up to subleading jet algorithm and NGL effects. Indeed, in Section \ref{sec:difference}, we defined the non-global contribution to the integrated jet thrust distribution by subtracting the refactorization ansatz in integrated form from our complete result for the scale-dependent part of the integrated distribution. If we had used a different subtraction procedure it is not clear that we would have been successful in making contact with the integrated hemisphere function. This remark is related to the fact that, at the outset of this work, it was not clear what appearance, if any, the jet cone size $R$ would make in the non-global contribution that remains once one subtracts off the integrated refactorization ansatz.

Our results suggest that the non-global contribution to the integrated jet thrust distribution at
small $R$ is simply
\be
\label{eq:ngexact}
K_{R\to 0}^{f,\,\,{\rm exact}}(\tom,\omega) = 2\,\mathcal{R}_f\left({\tom Q \over 2 R \omega}\right)\,,
\ee
where $\mathcal{R}_f(z)$ is the non-global contribution to the integrated hemisphere mass distribution.
We have only shown this up to a $\tom$- and $\omega$-independent function of $R$, due to the fact that we consistently dropped all contributions that depended on no scales at all or on $R$ but not on a ratio of dimensionful scales. 
This small $R$ limit depends on the precise definition of $K^f_R(\tom,\omega)$. In our definition, $K^f_R(\tom,\omega)$ is everything not given by the refactorization from~\cite{Kelley:2011tj}. With this definition, Eq.~\eqref{eq:ngexact} implies that all of the remaining singular $R$ dependence is inextricably linked to the ratio ${\tom Q \over 2 \omega}$. 

The assertion of the last paragraph is very sensible because we showed in Section \ref{sec:nglextract} that the leading and subleading non-global logarithms are
\begin{eqnarray}
\label{eq:asympt2}
K^{f,\,\text{NGL}}_{R \to 0}(\tom,\omega) &=& 
C_F C_A\Bigg[
-\frac{8\pi^2}{3}
\ln^2\left(\frac{\tau_\omega Q}{2 R \omega}\right)
+ \Bigg(
-\frac{8}{3} + \frac{88\pi^2}{9} -16\zeta_3\Bigg)
\left|\ln\left(\frac{\tau_\omega Q}{2 R \omega}\right)\right|
\Bigg]
\el
+ C_F n_f T_F
\Bigg(
\frac{16}{3} - \frac{32\pi^2}{9}\Bigg)
\left|\ln\left(\frac{\tau_\omega Q}{2 R \omega}\right)\right| + \cdots \,.
\end{eqnarray}
In the appropriate limit, Eq. \eqref{eq:ngexact} reduces to Eq. \eqref{eq:asympt2} up to terms not constrained by our calculation.
 The form of Eq. \eqref{eq:asympt2} also shows that the NGLs that arise in the integrated cone jet thrust distribution are naturally $R$ dependent. 
The great strength of our approach is that we abandoned the leading-logarithm approximation, preferring instead to first perform exact calculations and only then attempt to determine an appropriate approximation scheme.

The results recapitulated above motivated us to modify the refactorization formula of Ref.~\cite{Kelley:2011tj} to include the non-global contributions as well. In Section \ref{sec:nglextract}, we discussed the possibility of a refined refactorization formula
\begin{eqnarray}
\label{eq:refinedrefac2}
S_R(k_L,k_R,\lambda, \mu)
&=&
S_\mu\left(\frac{k_L}{\sqrt{R} \mu}\right)
S_\mu\left(\frac{k_R}{\sqrt{R} \mu}\right)
S_{\rm out}\left(\frac{2 \sqrt{R}\lambda}{\mu},R\right)
\el
~~~~~~~~~~~~~~~~~~~~~\otimes\,
S_f\left(\frac{k_L}{2 R\lambda}\right)
S_f\left(\frac{k_R}{2 R\lambda}\right)
+ \mathcal{O}(R)\,,
\end{eqnarray}
where the functions $S_\mu(z)$ and $S_f(z)$ denote respectively the $\mu$-dependent and $\mu$-independent parts of the hemisphere soft function~\cite{Chien:2010kc,Hoang:2008fs}. Eq. \eqref{eq:refinedrefac2} is consistent with all of the observations
made in this paper and, if true, would make precise the statement that resumming the NGLs in the cone jet thrust distribution is equivalent to resumming the NGLs in the hemisphere mass distribution.

In future work, it would be useful to understand to what extent refactorizations like Eq. \eqref{eq:refinedrefac2} 
hold beyond two loops. If this is understood, similar arguments should be appropriate for 
precision predictions of jet substructure observables at hadron colliders. In any case, it is of great interest to try and generalize Eq. \eqref{eq:refinedrefac2}, both to multi-jet processes at $e^+ e^-$ colliders and, if possible, to di- and multi-jet processes at hadron colliders. 

In order to understand non-global structure in more detail, it may be necessary to explore jet algorithms distinct from the thrust cone algorithm used in this paper. Most prominent among the alternatives is the anti-$k_T$ jet algorithm which, as noted in Section \ref{sec:smallr}, is very similar to the thrust cone one and has the advantage that it is widely used by experimentalists. It would
also be interesting to have all of the $R$-dependent constant terms that were dropped in writing down the non-global contribution to the integrated jet thrust distribution,  since they would significantly clarify the structure of the $\ln(R)$ terms that remain in the small $R$ limit.

Finally, even if Eq. \eqref{eq:refinedrefac2} turns out to be a useful approximation, one still has to deal with the hemisphere NGLs themselves. 
Resummation of these NGLs seems difficult~\cite{Dasgupta:2001sh,Banfi:2002hw,Kelley:2011ng,Marchesini:2003nh,Avsar:2009yb}.
Perhaps one can choose scales so that the non-global structure
has a numerically small effect. Nevertheless, it would be nice to understand NGLs in more detail, beyond the leading NGL.
One way forward would be to perform an explicit three-loop calculation of the integrated hemisphere soft function. While it goes without saying that this will be no easy task, we expect that the integrals which arise at three loops will turn out to be solvable using modern multi-loop computational techniques.

\section*{Acknowledgments}
RK and MDS were supported in part by the Department of Energy, under grant DE-SC003916.
HXZ was supported by the National Natural Science Foundation of China under grants No.~11021092 and No.~10975004. RMS gratefully acknowledges CICYT support through the project FPA-2009-09017 and CAM support through the project HEPHACOS S2009/ESP-1473. RMS also acknowledges the Research Executive Agency (REA) of the European Union under the Grant Agreement number PITN-GA-2010-264564 (LHCPhenoNet).

\appendix

\section{Calculation of the Integrated Jet Thrust Distribution}
\label{app:int}
In this appendix, we calculate the six terms in Eq. (\ref{eq:cumulative}) in the order in which we discussed the corresponding contributions to the NLO $\tau_\omega$ soft function in Section \ref{sec:calculation}. We begin with the charge renormalization contributions. The result is
\begin{eqnarray}
\label{eq:cumcharge}
&&K^{\rm Ren}_R(\tau_\omega,\omega,\mu) =
C_F C_A
\Bigg[
\frac{176}{9} \ln ^3\left(\frac{\mu }{\tau_\omega Q}\right)+\frac{88}{3} \ln (r) \ln ^2\left(\frac{\mu }{\tau_\omega Q}\right)+\left(\frac{44 \ln ^2(r)}{3}\right.
\el
\left.-\frac{22 \pi ^2}{9}\right)\ln \left(\frac{\mu
   }{\tau_\omega Q}\right)
+
\ln \left(\frac{\mu }{2 \omega }\right) \left(\frac{176 {\rm Li}_2(-r)}{3}+\frac{44 \ln ^2(r)}{3}+\frac{44 \pi ^2}{9}\right)-\frac{88}{3} \ln (r) \ln ^2\left(\frac{\mu }{2 \omega
   }\right)
\Bigg]
\el
+
C_F n_f T_F
\Bigg[
-\frac{64}{9} \ln ^3\left(\frac{\mu }{\tau_\omega Q}\right)-\frac{32}{3} \ln (r) \ln ^2\left(\frac{\mu }{\tau_\omega Q}\right)+\left(\frac{8 \pi ^2}{9}-\frac{16 \ln ^2(r)}{3}\right) \ln \left(\frac{\mu
   }{\tau_\omega Q}\right)
\el
+
\ln \left(\frac{\mu }{2 \omega }\right) \left(-\frac{64 {\rm Li}_2(-r)}{3}-\frac{16}{3} \ln ^2(r)-\frac{16 \pi ^2}{9}\right)+\frac{32}{3} \ln (r) \ln ^2\left(\frac{\mu }{2 \omega
   }\right)
\Bigg]\,.
\end{eqnarray}
For the real-virtual interference contributions, the result is
\begin{eqnarray}
  \label{eq:cumrr}
&&K^{\rm R-V}_R(\tau_\omega,\omega,\mu) =
C_F C_A
\Bigg[
\frac{64}{3} \ln ^4\left(\frac{\mu }{\tau_\omega Q}\right)+\frac{128}{3} \ln (r) \ln ^3\left(\frac{\mu }{\tau_\omega Q}\right)+\left(32 \ln ^2(r)\right.
\el
\left.-8 \pi ^2\right)
\ln ^2\left(\frac{\mu }{\tau_\omega Q}\right)+\ln
   \left(\frac{\mu }{\tau_\omega Q}\right) \left(\frac{32 \ln ^3(r)}{3}-8 \pi ^2 \ln (r)-\frac{64 \zeta (3)}{3}\right)
\el
+
\ln^2\left(\frac{\mu }{2 \omega }\right) \bigg(128 {\rm Li}_2(-r)+32 \ln ^2(r)+\frac{32 \pi ^2}{3}\bigg)+\ln \left(\frac{\mu }{2 \omega }\right)\left(128
   {\rm Li}_3\left(\frac{1}{r+1}\right)
\right.
\el
\left.
-128 {\rm Li}_3\left(\frac{r}{r+1}\right)-128 {\rm Li}_2(-r) \ln (r)+256 {\rm Li}_2(-r) \ln (r+1)-\frac{32}{3} \ln ^3(r)\right.
\el
\left.+64 \ln^2(r+1)
   \ln (r)+8 \pi ^2 \ln (r)+\frac{64}{3} \pi ^2 \ln (r+1)\right)-\frac{128}{3} \ln (r) \ln ^3\left(\frac{\mu }{2 \omega }\right)
\Bigg]
\end{eqnarray}
and for the same-side in-in contribution, we have
\begin{eqnarray}
  \label{eq:cumsdii}
&&K^{r_1}_R(\tau_\omega,\omega,\mu) =
C_F C_A
\Bigg[
-\frac{64}{3} \ln ^4\left(\frac{\mu }{\tau_\omega Q}\right)+\left(-\frac{128 \ln (r)}{3}-\frac{352}{9}\right) \ln ^3\left(\frac{\mu }{\tau_\omega Q}\right)
\el
+\left(-32 \ln ^2(r)-\frac{176 \ln
   (r)}{3}+\frac{16 \pi ^2}{3}-\frac{536}{9}\right) \ln ^2\left(\frac{\mu }{\tau_\omega Q}\right)+\ln \left(\frac{\mu }{\tau_\omega Q}\right) \left(-\frac{32}{3} \ln ^3(r)\right.
\el
\left.-\frac{88 \ln^2(r)}{3}+\frac{16}{3} \pi ^2 \ln (r)-\frac{536 \ln (r)}{9}+\frac{232 \zeta (3)}{3}-\frac{22 \pi ^2}{9}-\frac{1544}{27}\right)
\Bigg]
\el
+
C_F n_f T_F
\Bigg[
\frac{128}{9} \ln ^3\left(\frac{\mu }{\tau_\omega Q}\right)+\left(\frac{64 \ln (r)}{3}+\frac{160}{9}\right)\ln^2\left(\frac{\mu }{\tau_\omega Q}\right)+\left(\frac{32 \ln ^2(r)}{3}+\frac{160 \ln(r)}{9}\right.
\el
\left.+\frac{8 \pi ^2}{9}+\frac{304}{27}\right) \ln \left(\frac{\mu }{\tau_\omega Q}\right)
\Bigg]\,.
\end{eqnarray}

The calculation of the opposite-side in-in contributions is somewhat less trivial since the integral over $S^{r_2}_R(k_L,k_R,\lambda,\mu)$ requires sector decomposition. Carrying out this analysis leads to
\cmb{-1 cm}{0 cm}
\bea
&&C_F C_A
\Bigg[-2\ln \left(\frac{\mu }{\tau_\omega Q}\right)\int_0^1  {\dd z\over z}\left({2\mbox{\Large $f$}_{C_A}^{(0)}\left({z \over 2-z},r\right)\over 2-z} - \mbox{\Large $f$}_{C_A}^{(0)}\left(0,r\right)\right) 
\el + 2 \ln (2)\mbox{\Large $f$}_{C_A}^{(0)}\left(0,r\right)\ln \left(\frac{\mu }{\tau_\omega Q}\right) + \mbox{\Large $f$}_{C_A}^{(1)}\left(0,r\right)\ln \left(\frac{\mu }{\tau_\omega Q}\right)+2 \mbox{\Large $f$}_{C_A}^{(0)}\left(0,r\right)\ln^2 \left(\frac{\mu }{\tau_\omega Q}\right)\Bigg]
\el
+C_F n_f T_F\Bigg[-2\ln \left(\frac{\mu }{\tau_\omega Q}\right)\int_0^1  {\dd z\over z}\left({2\mbox{\Large $f$}_{n_f}^{(0)}\left({z \over 2-z},r\right)\over 2-z}- \mbox{\Large $f$}_{n_f}^{(0)}\left(0,r\right)\right)
\el
 + 2 \ln (2)\mbox{\Large $f$}_{n_f}^{(0)}\left(0,r\right)\ln \left(\frac{\mu }{\tau_\omega Q}\right)+\mbox{\Large $f$}_{n_f}^{(1)}\left(0,r\right)\ln \left(\frac{\mu }{\tau_\omega Q}\right)+2 \mbox{\Large $f$}_{n_f}^{(0)}\left(0,r\right)\ln^2 \left(\frac{\mu }{\tau_\omega Q}\right)\Bigg]
\label{secdecompinin}
\eea
\cme
for the finite part of the scale-dependent terms. Eq. (\ref{secdecompinin}) is written in terms of the functions introduced in Section \ref{sec:calculation}. We can easily evaluate the above expression using the moments tabulated in Eqs. (\ref{momentsinin}) and the result is
\cmb{-.7 cm}{0 cm}
\begin{eqnarray}
\label{eq:cumosii}
&&K^{r_2}_R(\tau_\omega,\omega,\mu) =
C_F C_A
\Bigg[  
\left(\frac{176 {\rm Li}_2\left(r^2\right)}{3}+32 {\rm Li}_3\left(\frac{r^2}{r^2-1}\right)-32 {\rm Li}_2\left(r^2\right) \ln \left(1-r^2\right)
\right.
\el
\left.
+24 {\rm Li}_2\left(r^2\right) \ln
   \left(r^2\right)+\frac{16 r^2}{3-3 r^2}-\frac{16}{3} \ln ^3\left(1-r^2\right)+\frac{88}{3} \ln \left(r^2\right) \ln \left(1-r^2\right)\right.
\el
\left.+\frac{32 r^2 \ln (r)}{3
   \left(r^2-1\right)^2}\right) \ln \left(\frac{\mu }{\tau_\omega Q}\right)+(64 {\rm Li}_2(-r)+64 {\rm Li}_2(r)) \ln ^2\left(\frac{\mu }{\tau_\omega Q}\right)
\Bigg]+
C_F n_f T_F
\Bigg[
\left(-\frac{64 {\rm Li}_2\left(r^2\right)}{3}\right.
\el
\left.
+\frac{32 r^2}{3 \left(r^2-1\right)}-\frac{64 r^2 \ln (r)}{3 \left(r^2-1\right)^2}-\frac{64}{3} \ln (r) \ln \left(1-r^2\right)\right) \ln \left(\frac{\mu }{\tau_\omega Q}\right)
\Bigg]\,.
\end{eqnarray}
\cme

The calculation of the in-out contributions requires sector decomposition as well. In terms of the functions introduced in Section \ref{sec:calculation}, the scale-dependent terms are given by
\bea
&&C_F C_A
\Bigg[ \mbox{\Large $g$}_{C_A}^{(1)}\left(0,r\right)\ln \left(\frac{\mu }{2\omega}\right) + \mbox{\Large $g$}_{C_A}^{(0)}\left(0,r\right)\ln^2 \left(\frac{\mu}{2\omega}\right)+\mbox{\Large $g$}_{C_A}^{(0)}\left(0,r\right)\ln^2\left(\frac{\mu }{\tau_\omega Q}\right)
\el
-\ln\left(\frac{\mu }{2\omega}\right)\int_0^1{\dd z\over z}\left(\mbox{\Large $g$}_{C_A}^{(0)}\left(z,r\right) + \mbox{\Large $g$}_{C_A}^{(0)}\left(1/z,r\right) - 2 \mbox{\Large $g$}_{C_A}^{(0)}\left(0,r\right)\right) - \frac{1}{2}\mbox{\Large $g$}_{C_A}^{(1)}\left(0,r\right)\ln \left(\frac{\tau_\omega Q}{2\omega}\right)
\el
 - \frac{1}{2}\mbox{\Large $g$}_{C_A}^{(0)}\left(0,r\right)\ln^2 \left(\frac{\tau_\omega Q}{2\omega}\right)+\int_1^{\tau_\omega Q \over 2 \omega} {\dd z_2 \over z_2}\int_0^1  {\dd z_1 \over z_1}\left(\mbox{\Large $g$}_{C_A}^{(0)}\left({z_2 \over z_1},r\right) - \mbox{\Large $g$}_{C_A}^{(0)}\left(0,r\right)\right)\Bigg]
\el
+
C_F n_f T_F
\Bigg[ \mbox{\Large $g$}_{n_f}^{(1)}\left(0,r\right)\ln \left(\frac{\mu }{2\omega}\right) + \mbox{\Large $g$}_{n_f}^{(0)}\left(0,r\right)\ln^2 \left(\frac{\mu}{2\omega}\right)+\mbox{\Large $g$}_{n_f}^{(0)}\left(0,r\right)\ln^2\left(\frac{\mu }{\tau_\omega Q}\right)
\el
-\ln\left(\frac{\mu }{2\omega}\right)\int_0^1{\dd z\over z}\left(\mbox{\Large $g$}_{n_f}^{(0)}\left(z,r\right) + \mbox{\Large $g$}_{n_f}^{(0)}\left(1/z,r\right) - 2 \mbox{\Large $g$}_{n_f}^{(0)}\left(0,r\right)\right) - \frac{1}{2}\mbox{\Large $g$}_{n_f}^{(1)}\left(0,r\right)\ln \left(\frac{\tau_\omega Q}{2\omega}\right)
\el
 - \frac{1}{2}\mbox{\Large $g$}_{n_f}^{(0)}\left(0,r\right)\ln^2 \left(\frac{\tau_\omega Q}{2\omega}\right)+\int_1^{\tau_\omega Q \over 2 \omega} {\dd z_2 \over z_2}\int_0^1  {\dd z_1 \over z_1}\left(\mbox{\Large $g$}_{n_f}^{(0)}\left({z_2 \over z_1},r\right) - \mbox{\Large $g$}_{n_f}^{(0)}\left(0,r\right)\right)\Bigg]\,.
\label{secdecinout}
\eea
Using the moments tabulated in Eqs. (\ref{momentsinout}), we find an explicit expression for  Eq. (\ref{secdecinout}):
\cmb{-.5 cm}{0 cm}
\begin{eqnarray}
\label{eq:cumio}
&&K^{r_3}_R(\tau_\omega,\omega,\mu) =
C_F C_A
\Bigg[
\left(-64 {\rm Li}_2(-r)-64 {\rm Li}_2(r)+\frac{16 \pi^2}{3}\right) \ln^2\left(\frac{\mu }{\tau_\omega Q}\right) 
\el
+\left(-64 {\rm Li}_2(-r)-64 {\rm Li}_2(r)+\frac{16 \pi ^2}{3}\right) \ln ^2\left(\frac{\mu }{2 \omega }\right)+\ln \left(\frac{\mu }{2 \omega }\right)\left(-\frac{352
   {\rm Li}_2\left(r^2\right)}{3}
\right.
\el
\left.
+128 {\rm Li}_3(1-r)+64 {\rm Li}_3(r)-64 {\rm Li}_3\left(\frac{1}{r+1}\right)+64 {\rm Li}_3\left(\frac{r}{r+1}\right)-128 {\rm Li}_2(-r) \ln
   (r+1)
\right.
\el
\left.
+128 {\rm Li}_2(r) \ln (1-r)-64 {\rm Li}_2(r) \ln (r)+\frac{16 \left(r^2+1\right)}{3 \left(r^2-1\right)}-\frac{64 r^2 \ln (r)}{3 \left(r^2-1\right)^2}-32 \ln (r) \ln
   ^2(r+1)
\right.
\el
\left.
+64 \ln ^2(1-r) \ln (r)-\frac{64}{3} \pi ^2 \ln (1-r)-\frac{352}{3} \ln (1-r) \ln (r)-\frac{352}{3} \ln (r) \ln (r+1)\right.
\el 
\left.-\frac{32}{3} \pi ^2 \ln (r+1)-64 \zeta_3 +\frac{176 \pi ^2}{9}\right)+\ln \left(\frac{\tau _{\omega}Q}{2 \omega
   }\right)\left(8 {\rm Li}_3\left(r^2\right)-64
   {\rm Li}_3\left(\frac{r}{r+1}\right)+48 \zeta_3\right.
\el
\left.-64 {\rm Li}_3(1-r)+64 {\rm Li}_2(-r) \ln (r+1)-32
   {\rm Li}_2(-r) \ln (r)-64 {\rm Li}_2(r) \ln (1-r)+\frac{32}{3} \ln
   ^3(r+1)\right.
\el
\left.+\frac{32}{3} \pi ^2 \ln (1-r)-32 \ln ^2(1-r) \ln (r)\right)+\ln^2\left(\frac{\tau _{\omega}Q}{2 \omega}\right)\bigg(32 {\rm Li}_2(-r)-\frac{8 \pi^2}{3}+32 {\rm Li}_2(r)\bigg)
\el
 +
\mbox{\large $\chi$}_{C_A}\left(\frac{\tau_\omega Q}{2\omega},r\right)-\mbox{\large $\chi$}_{C_A}\left(1,r\right)
\Bigg]+ C_F n_f T_F
\Bigg[
\ln \left(\frac{\mu }{2 \omega }\right) \left(\frac{128 {\rm Li}_2\left(r^2\right)}{3}-\frac{64 \pi ^2}{9}-\frac{32 \left(r^2+1\right)}{3
   \left(r^2-1\right)}\right.
\el
\left.+\frac{128 r^2 \ln (r)}{3 \left(r^2-1\right)^2}+\frac{128}{3} \ln (r) \ln \left(1-r^2\right)\right)+\mbox{\large $\chi$}_{n_f}\left(\frac{\tau_\omega Q}{2\omega},r\right)-\mbox{\large $\chi$}_{n_f}\left(1,r\right)
\Bigg]\,,
\end{eqnarray}
\cme
where $\mbox{\large $\chi$}_{C_A}\left(x,r\right)$ and $\mbox{\large $\chi$}_{n_f}\left(x,r\right)$ are non-trivial functions built out of one- and two-dimensional harmonic polylogarithms. They are given explicitly in Appendix \ref{sec:anacumulant}.

Finally, for the out-out contribution, the result is
\begin{eqnarray}
\label{eq:cumoo}
&&K^{r_4}_R(\tau_\omega,\omega,\mu) =
C_F C_A
\Bigg[
\ln \left(\frac{\mu }{2 \omega }\right) \left(-32 {\rm Li}_3\left(\frac{r^2}{r^2-1}\right)+\frac{352 {\rm Li}_2(r)}{3}
\right.
\el
\left.
-64 {\rm Li}_3(r)-64
   {\rm Li}_3\left(\frac{1}{r+1}\right)+64 {\rm Li}_3\left(\frac{r}{r+1}\right)+64 {\rm Li}_2(-r) \ln (1-r)
\right.
\el
\left.
+32 {\rm Li}_2(-r) \ln (r)-64
   {\rm Li}_2(-r) \ln (r+1)-32 {\rm Li}_2(r) \ln (r)+64 {\rm Li}_2(r) \ln (r+1)
\right.
\el
\left.
+\frac{32 r^2 \ln (r)}{3 \left(r^2-1\right)^2}
+\frac{16}{3} \ln
   ^3(1-r)+\frac{32 \ln ^3(r)}{3}+\frac{16}{3} \ln ^3(r+1)-64 \ln (r) \ln ^2(1-r)
\right.
\el
\left.
+16 \ln (r+1) \ln ^2(1-r)+16 \ln ^2(r+1) \ln (1-r)-\frac{88 \ln ^2(r)}{3}
-32 \ln (r) \ln
   ^2(r+1)
\right.
\el
\left.
+\frac{176}{3} \ln (r) \ln (1-r)+\frac{64}{3} \pi ^2 \ln (1-r)-\frac{16}{3} \pi ^2 \ln (r)+\frac{536 \ln (r)}{9}-\frac{32}{3} \pi ^2+64 \zeta_3
\right.
\el
\left.
-\frac{176 \pi ^2}{9}+\frac{176}{3} \ln (r) \ln (r+1) \ln (r+1)-128 {\rm Li}_3(1-r)+\frac{8 r^2+8}{3-3 r^2}-64 {\rm Li}_2(r) \ln (1-r)\right)
\el
+\ln ^2\left(\frac{\mu }{2 \omega }\right) \bigg(-64 {\rm Li}_2(-r)+64 {\rm Li}_2(r)-32 \ln ^2(r)+\frac{176 \ln (r)}{3}-16
   \pi ^2\bigg)
\el
+\frac{128}{3} \ln (r) \ln ^3\left(\frac{\mu }{2 \omega }\right)  
\Bigg]
\el
+ C_F n_f T_F
\Bigg[
\ln \left(\frac{\mu }{2 \omega }\right) \left(-\frac{128 {\rm Li}_2(r)}{3}-\frac{16 \left(r^2+1\right)}{3 \left(1-r^2\right)}-\frac{64 r^2 \ln (r)}{3 \left(1-r^2\right)^2}+\frac{32
   \ln ^2(r)}{3}-\frac{160 \ln (r)}{9}
\right.
\el
\left.
-\frac{64}{3} \ln (r) \ln \left(\frac{1-r}{r+1}\right)-\frac{128}{3} \ln (r) \ln (r+1)+\frac{64 \pi ^2}{9}\right)
-\frac{64}{3} \ln (r) \ln
   ^2\left(\frac{\mu }{2 \omega }\right)
\Bigg]\,.
\end{eqnarray}
Now that all the pieces are in place, we can combine them together and study the result.

\section{Analytic Expressions For Non-Trivial Two-Parameter Integrals}
\label{app:chi}
\label{sec:anacumulant}
The non-trivial two-parameter integrals
\bea
&&\int_1^{x} {\dd z_2 \over z_2}\int_0^1  {\dd z_1\over z_1}\left(\mbox{\Large $g$}_{C_A}^{(0)}\left({z_2\over z_1},r\right) - \mbox{\Large $g$}_{C_A}^{(0)}\left(0,r\right)\right)
\el
\int_1^{x} {\dd z_2 \over z_2}\int_0^1  {\dd z_1\over z_1}\left(\mbox{\Large $g$}_{n_f}^{(0)}\left({z_2\over z_1},r\right) - \mbox{\Large $g$}_{n_f}^{(0)}\left(0,r\right)\right)\,,
\eea
first encountered in Section \ref{sec:calculation}, were parametrized there in terms of two functions, $\mbox{\Large $\chi$}_{C_A}(x,r)$ and $\mbox{\Large $\chi$}_{n_f}(x,r)$, such that:
\bea
&&\int_1^{x} {\dd z_2 \over z_2}\int_0^1  {\dd z_1\over z_1}\left(\mbox{\Large $g$}_{C_A}^{(0)}\left({z_2\over z_1},r\right) - \mbox{\Large $g$}_{C_A}^{(0)}\left(0,r\right)\right) = \mbox{\Large $\chi$}_{C_A}\left(x,r\right) - \mbox{\Large $\chi$}_{C_A}\left(1,r\right)
 \el
\int_1^{x} {\dd z_2 \over z_2}\int_0^1  {\dd z_1\over z_1}\left(\mbox{\Large $g$}_{n_f}^{(0)}\left({z_2\over z_1},r\right) - \mbox{\Large $g$}_{n_f}^{(0)}\left(0,r\right)\right) = \mbox{\Large $\chi$}_{n_f}\left(x,r\right) - \mbox{\Large $\chi$}_{n_f}\left(1,r\right)\,.
\eea
As explained in Section \ref{sec:calculation}, $\mbox{\Large $g$}_{C_A}^{(0)}(x,r)$ and $\mbox{\Large $g$}_{n_f}^{(0)}(x,r)$ are simply the leading order terms in the epsilon expansion of the in-out contributions to the $\tau_\omega$ soft function. In this appendix we provide analytical expressions for both $\mbox{\Large $\chi$}_{C_A}(x,r)$ and $\mbox{\Large $\chi$}_{n_f}(x,r)$. In what follows the $H$ functions are one- or two-dimensional harmonic polylogarithms (introduced in Refs. \cite{Remiddi:1999ew} and \cite{Gehrmann:2000zt} respectively).

For the $C_F C_A$ color structure we have
\cmb{-.4 cm}{0 cm}
\bea
&&\mbox{\Large $\chi$}_{C_A}(x,r) = -\frac{32 r^2 (x-1)}{3 (2 r+1) (x r+r+x)}+\frac{32 \left(2 H(0,r)+H\left(\frac{r^2}{r+1},x\right)\right) x^3}{3
   (x+1) (x r+r+x)}\\
&&-\frac{4 \left(H(-1,r)-5 H(0,r)+H\left(0,1-r^2\right)+4
   H(0,x)-4 H\left(\frac{r}{r+1},x\right)\right) x}{3 (x+1)}
\el
+\frac{16 (1-2 x) H\left(\frac{1}{r+1},x\right) x}{3 (x
   r+r+x)}+\frac{16 \left(2
   H(0,r)+H\left(\frac{r^2}{r+1},x\right)\right) x^3}{3 (x+1)^2 (x
   r+r+x)}
\el
-\frac{16 (r+1) x}{3 (x+1)
   (x r+r+x)}\bigg(H(-1,r)-H(0,r)+H\left(0,1-r^2\right)+H(0,x)\bigg)
\el
+\frac{32 x}{3 (r+1)}\bigg(-2
   H(0,r)+H\left(\frac{1}{r+1},x\right)-H\left(\frac{r^2}{r+1},x\right)
   \bigg)-\frac{32
   \left(H(0,x)-H\left(\frac{1}{r+1},x\right)\right)}{3
   \left(r^2-1\right)}
\el
+\frac{4 \left(3 H(-1,r)+5 H(0,r)+3
   H\left(0,1-r^2\right)+8 H\left(\frac{r}{r+1},x\right)\right)}{3
   (x+1)}+\frac{16}{3 (r+1)^2} \bigg(-2
   H(0,r)
\el
+H\left(\frac{1}{r+1},x\right)-H\left(\frac{r^2}{r+1},x\right)
   \bigg)-\frac{16}{3 (x+1) (r x+x+1)}\bigg(H(-1,r)-2 H(0,r)
\el
+H\left(0,1-r^2\right)+H(0,x)-H\left(\frac{r^2}{r+1},x\right)\bigg)-\frac{32}{3 (r+1)}\left(-H(0,r)+H\left(\frac{r}{r+1},x\right)\right.
\el
\left.-H\left(\frac{r^2}{r+1},x\right)\right)-\frac{16 (x+2) \left(2
   H(0,r)+H\left(\frac{r^2}{r+1},x\right)\right)}{3 (x+1)^2}-\frac{8 r }{3 (r+1)^2}\left(2 H(0,r) \bigg(H(0,x)\right.
\el
\left.-4
   H\left(\frac{r}{r+1},x\right)\bigg)-H\left(0,\frac{1}{r+1},x\right)
   +H\left(0,\frac{r^2}{r+1},x\right)+4
   H\left(\frac{r}{r+1},\frac{1}{r+1},x\right)\right.
\el
\left.-4
   H\left(\frac{r}{r+1},\frac{r^2}{r+1},x\right)\right)+\frac{16 x}{3
   (x+1)^2} \left(2 H(0,r) H(0,x)-
   \left(H\left(\frac{1}{r+1},x\right)\right.\right.
\el
\left.\left.-H\left(\frac{r}{r+1},x\right)\right) H(0,x)-\bigg(H(-1,r)-2 H(0,r) + H\left(0,1-r^2\right)\bigg)
   H\left(\frac{1}{r+1},x\right)\right.
\el
\left.+\bigg(H(-1,r)-3 H(0,r)+
   H\left(0,1-r^2\right)\bigg) H\left(\frac{r}{r+1},x\right)+2
   H(0,-1,r)-5 H(0,0,r)\right.
\el
\left.- H\left(0,1,r^2\right)+
   H\left(0,\frac{1}{r+1},x\right)-2 H\left(0,\frac{r}{r+1},x\right)+
   H\left(0,\frac{r^2}{r+1},x\right)\right.
\el
\left.+
   H\left(\frac{1}{r+1},\frac{r^2}{r+1},x\right)- H\left(\frac{r}{r+1},\frac{r^2}{r+1},x\right)\right)+\frac{8 r}{3 (r-1)^2} \bigg(2 H(0,r)
   H(0,x)
\el
-H\left(0,\frac{1}{r+1},x\right)+H\left(0,\frac{r^2}{r+1},x\right)\bigg)+\frac{176}{3} H(-1,x) \bigg(2 H(0,-1,r)-5 H(0,0,r)
\el
-H\left(0,1,r^2\right)\bigg)+\frac{176}{3} H(0,x) \bigg(2 H(0,r) H\left(0,1-r^2\right)+2 H\left(0,1,r^2\right)-\frac{\pi^2}{3}\bigg)
\el
+\frac{176}{3}
   \left(-H(-1,x) H(0,x) H\left(\frac{1}{r+1},x\right)+\bigg(H(-1,x)+H(0,x)\bigg)
   H\left(0,\frac{1}{r+1},x\right)\right.
\el
\left.-\bigg(2 H(-1,x)+H(0,x)\bigg)H\left(0,\frac{r}{r+1},x\right)+H(0,x)\left(H\left(-1,\frac{r}{r+1},x\right)\right.\right.
\el
\left.\left.+H\left(\frac{1}{r+1},-1,x\right)\right)+H\left(-1,0,\frac{r^2}{r+1},x\right)+H\left(-1, \frac{1}{r+1},\frac{r^2}{r+1},x\right)
\right.
\el
\left.-H\left(-1,\frac{r}{r+1},\frac{r^2}{r+1},x\right)+H\left(0,-1,\frac{r}{r+1},x\right)-3 H\left(0,0,\frac{1}{r+1},x\right)\right.
\el
\left.+3 H\left(0,0,\frac{r}{r+1},x\right)-H\left(0,\frac{1}{r+1},-1,x\right) -H\left(0,\frac{1}{r+1},\frac{r^2}{r+1},x\right)\right.
\el
\left.+2 H\left(0,\frac{r}{r+1},-1,x\right)+ H\left(0,\frac{r}{r+1},\frac{1}{r+1},x\right)\right) +\frac{176}{3}\bigg(H(-1,r)
\el
+H\left(0,1-r^2\right)\bigg)\left(-H\left(-1,\frac{1}{r+1},x\right)+H\left(-1,\frac{r}{r+1},x\right)+H\left(0,\frac{1}{r+1},x\right)\right.
\el
\left.-H\left(0,\frac{r}{r+1},x\right)\right)+\frac{176}{3} H(0,r) \left(2 H(-1,0,x)+2
   H\left(-1,\frac{1}{r+1},x\right)\right.
\el
\left.-3
   H\left(-1,\frac{r}{r+1},x\right)-2 H\left(0,\frac{1}{r+1},x\right)+H\left(0,\frac{r}{r+1},x\right)\right)+ 8 H(0,-1,x) \bigg(\frac{\pi^2}{3}
\el
-4 H(0,-1,r)+10 H(0,0,r)\bigg)+16 H(-1,r)\left(H\left(0,-1,\frac{1}{r+1},x\right)\right.
\el
\left.-H\left(0,-1,\frac{r}{r+1},x
   \right)-H\left(0,0,\frac{1}{r+1},x\right)+H\left(0,0,\frac{r}{r+1},x
   \right)\right)
\el
-16 H(0,r) \left(2 H(0,-1,0,x)+2 H\left(0,-1,\frac{1}{r+1},x\right)-3 H\left(0,-1,\frac{r}{r+1},x\right)\right.
\el
\left.- 2 H\left(0,0,-1,\frac{x}{r}\right) - 2 H\left(0,0,\frac{1}{r+1},x\right)+3 H\left(0,0,\frac{r}{r+1},x\right)\right)
\el
+16 \left(H(0,x) \bigg(H\left(0,-1,\frac{1}{r+1},x\right)-
   H\left(0,0,\frac{1}{r+1},x\right)+H\left(0,0,\frac{r}{r+1},x\right)\bigg)\right.
\el
\left.-H\left(0,-1,0,\frac{r^2}{r+1},x\right)-H\left(0,-1,\frac{1}{r+1},\frac{r^2}{r+1},x\right)-H  \left(0,-1,
   \frac{r}{r+1},0,x\right)\right.
\el
\left.+H\left(0,-1,\frac{r}{r+1},\frac{r}{r+1},x\right)-H\left
   (0,-1,\frac{r}{r+1},\frac{1}{r+1},x\right)+H\left(0,-1,\frac{r}{r+1},\frac{r^2}{r+1},x\right)\right.
\el
\left.-2
   H\left(0,0,-1,\frac{1}{r+1},x\right)-H\left(0,0,-1,\frac{1}{r+1},\frac{x}{r}\right)+ H\left(0,0,-1,\frac{r}{r+1},\frac{x}{r}\right) \right.
\el
\left.+ 2 H\left(0,0,0,\frac{1}{r+1},x\right)-2 H\left(0,0,0,\frac{r}{r+1},x\right)+H\left(0,0,\frac{1}{r+1},\frac{r^2}{r+1},x\right)\right.
\el
\left.-H\left(0,0,\frac{r}{r+1},\frac{r}{r+1},x\right)+H\left(0,0,\frac{r}{r+1},\frac{1}{r+1},x\right)- H\left(0,0,\frac{r}{r+1},\frac{r^2}{r+1},x\right)\right)
\el
+32 H(0,x) \bigg(H(0,r)H\left(0,1,r^2\right)-H\left(0,0,1,r^2\right)+\zeta_3\bigg)
\nonumber
\eea
and for the $C_F n_f T_F$ color structure we have
\bea
&&\mbox{\Large $\chi$}_{n_f}(x,r) = -\frac{64 (1-x) r^2}{3 (2 r+1) (x r+r+x)}+\frac{32 (r+1) r}{3 (x r+r+x)}
   \left(-H(-1,r)+H(0,r)\right.
\\
&&\left.-H\left(0,1-r^2\right)-H(0,x)\right)+\frac{32 H\left(\frac{1}{r+1},x\right) r}{(r+1) (x r+r+x)}-\frac{64 H\left(\frac{1}{r+1},x\right) r}{3 (r+1)^2 (x r+r+x)}
\el
+\frac{32 \left(2
   H(0,r)-H\left(\frac{1}{r+1},x\right)+H\left(\frac{r^2}{r+1},x\right)\right) r}{3 (r+1)^2}-\frac{32 (r-1) \left(H(0,r)+H\left(\frac{r}{r+1},x\right)\right)}{3 (r+1)}
\el
+\frac{64 H\left(\frac{1}{r+1},x\right) r}{3 \left(1-r^2\right)}-\frac{32
   \left(-H(0,r)+H\left(\frac{r}{r+1},x\right)-H\left(\frac{r^2}{r+1},x\right)\right)}{3 (x+1)}-\frac{32 H(0,r) r}{3 (x+1)}
\el
+\frac{32 \left(r^2+1\right) H(0,x)}{3 \left(r^2-1\right)}- \frac{64 \left(2 H(0,r)+H\left(\frac{r^2}{r+1},x\right)\right) r^4}{3 (r+1)^2 (x r+r+x)}\el
+\frac{32 \left(r^2-1\right) \left(H(-1,r)-2
   H(0,r)+H\left(0,1-r^2\right)+H(0,x)-H\left(\frac{r^2}{r+1},x\right)\right)}{3 (x+1) r}
\el
+\frac{32 (r+1) \left(H(-1,r)-2
   H(0,r)+H\left(0,1-r^2\right)+H(0,x)-H\left(\frac{r^2}{r+1},x\right)\right)}{3 (r x+x+1) r}
\el
+ \frac{32 \left(2 H(0,r)+H\left(\frac{r^2}{r+1},x\right)\right) r^3}{(r+1)(x r+r+x)}-\frac{16 (H(-1,r) H(0,x)+H(0,0,x)) r}{3 (r-1)^2}
\el
-\frac{64 \left((2 H(0,r)-H(-1,r)) H(0,x)-H(0,0,x)-H\left(0,\frac{1}{r+1},x\right)+H\left(0,\frac{r^2}{r+1},x\right)\right) r^2}{3\left(r^2-1\right)^2}
\el
-\frac{32 x}{3 (x+1)^2} \left(H(0,x) \bigg(H\left(\frac{r}{r+1},x\right)- H\left(\frac{1}{r+1},x\right)\bigg)+H(0,r) \bigg( 2 H(0,x)\right.
\el
\left.-H\left(\frac{r}{r+1},x\right)\bigg)+\bigg(H(-1,r)-2 H(0,r)+H\left(0,1-r^2\right)\bigg)\left(H\left(\frac{r}{r+1},x\right)\right.\right.
\el
\left.\left.-H\left(\frac{1}{r+1},x\right)\right)-5 H(0,0,r)+2 H(0,1,r)-2 H\left(0,1,r^2\right)+H\left(0,\frac{1}{r+1},x\right)\right.
\el
\left.-2 H\left(0,\frac{r}{r+1},x\right) +H\left(0,\frac{r^2}{r+1},x\right)+H\left(\frac{1}{r+1},\frac{r^2}{r+1},x\right)-H\left(\frac{r}{r+1},\frac{r^2}{r+1},x\right)\right)
\el
+\frac{16 r}{3(r+1)^2} \left(H(-1,r)
   H(0,x)-8 H(0,r) H\left(\frac{r}{r+1},x\right)+H(0,0,x)\right.
\el
\left.+4 \bigg(H\left(\frac{r}{r+1},\frac{1}{r+1},x\right)- H\left(\frac{r}{r+1},\frac{r^2}{r+1},x\right)\bigg)\right)+\frac{128}{3} H(0,x) \left(-H\left(0,1,r^2\right)\right.
\el
\left.+\frac{\pi ^2}{6}-H(0,r)
   H\left(0,1-r^2\right)\right)+\frac{64}{3} H(-1,x) \left(-2 H(0,-1,r)+5 H(0,0,r)\right.
\el
\left.+H\left(0,1,r^2\right)\right)+\frac{64}{3}
   H(0,r) \left(-H\left(-1,\frac{1}{r+1},x\right)+2 \bigg(H\left(-1,\frac{r}{r+1},x\right)\right.
\el
\left.-H(-1,0,x)\bigg)+H\left(0,\frac{1}{r+1},x\right)\right)+\frac{64}{3}H(0,x)\left(H\left(-1,\frac{1}{r+1},x\right)\right.
\el
\left.-H\left(-1,\frac{r}{r+1},x\right)-H\left(0,\frac{1}{r+1},x\right)+H\left(0,\frac{r}{r+1},x\right)\right)+\frac{64}{3}
\bigg(H(-1,r)
\el
-H(0,r)+H\left(0,1-r^2\right)\bigg)\left(H\left(-1,\frac{1}{r+1},x\right)-H\left(-1,\frac{r}{r+1},x\right)-H\left(0,\frac{1}{r+1},x\right)\right.
\el
\left.+H\left(0,\frac{r}{r+1},x\right)\right)+\frac{64}{3}H(-1,x) \left(2 H\left(0,\frac{r}{r+1},x\right) - H\left(0,\frac{1}{r+1},x\right)\right) 
\el
-\frac{64}{3}\left(H\left(-1,0,\frac{r^2}{r+1},x\right)+H\left(-1,\frac{1}{r+1},\frac{r^2}{r+1},x\right)-H\left(-1,\frac{r}{r+1},\frac{r^2}{r+1},x\right)\right.
\el
\left.+H\left(0,-1,\frac{r}{r+1},x\right)-3 H\left(0,0,\frac{1}{r+1},x\right) + 3 H\left(0,0,\frac{r}{r+1},x\right)-H\left(0,\frac{1}{r+1},-1,x\right)\right.
\el
\left.-H\left(0,\frac{1}{r+1},\frac{r^2}{r+1},x\right)+2
   H\left(0,\frac{r}{r+1},-1,x\right)+H\left(0,\frac{r}{r+1},\frac{1}{r+1},x\right)\right)\,.
\nonumber
\eea
\cme
\bibliographystyle{JHEP3}
\bibliography{tauomega}
\end{document}